\documentclass[A4paper, reqno,12pt]{amsart}

\usepackage{amsmath,amsthm,amssymb,color,graphicx,epsfig}
\usepackage{amscd,verbatim}
\usepackage[all]{xy}

\theoremstyle{plain}
\newtheorem{theorem}{Theorem}
\newtheorem{proposition}[theorem]{Proposition}

\newtheorem{corollary}[theorem]{Corollary}

\theoremstyle{definition}
\newtheorem{definition}[theorem]{Definition}

\theoremstyle{remark}

\newtheorem*{example}{Example}

\numberwithin{equation}{section}
\numberwithin{theorem}{section}
\numberwithin{figure}{section}
\numberwithin{table}{section}

\newcommand{\cA}{{\mathcal A}}
\newcommand{\cB}{{\mathcal B}}
\newcommand{\cC}{{\mathcal C}}

\newcommand{\cH}{{\mathcal H}}

\newcommand{\cK}{{\mathcal K}}

\newcommand{\cX}{{\mathcal X}}

\newcommand{\alg}{{\cA}}
\newcommand{\cb}{{\cB}}
\newcommand{\cc}{{\cC}}
\newcommand{\hilb}{{\cH}}
\newcommand{\cx}{{\cX}}
\newcommand{\cpt}{{\cK}}

\newcommand{\CC}{{\mathbb C}}

\newcommand{\RR}{{\mathbb R}}
\newcommand{\TT}{{\mathbb T}}
\newcommand{\ZZ}{{\mathbb Z}}
\newcommand{\torus}{\TT}
\newcommand{\integer}{\ZZ}
\newcommand{\real}{\RR}
\newcommand{\complex}{\CC}

\newcommand{\wt}[1]{\widetilde{#1}}
\newcommand{\wh}[1]{\widehat{#1}}
\newcommand{\uind}{u\hbox{-\rm{ind}}}
\newcommand{\urind}{u_{\rho}\hbox{-\rm{ind}}}
\newcommand{\conj}[1]{\overline{#1}}

\newcommand{\ft}{{\mathfrak t}}

\newcommand{\sfG}{{\mathsf G}}
\newcommand{\sfH}{{\mathsf H}}
\newcommand{\sfT}{{\mathsf T}}
\newcommand{\sfA}{{\mathsf A}}
\newcommand{\sfN}{{\mathsf N}}

\newcommand{\ad}{{\rm{ad}}}
\newcommand{\inn}{{\rm{Inn}}}
\newcommand{\aut}{{\rm{Aut}}}
\newcommand{\out}{{\rm{Out}}}

\newcommand{\ind}{{\rm{ind}}}

\begin{document}

\title[Nonassociative tori and applications to T-duality]{Nonassociative 
tori and applications to T-duality}

\author[P Bouwknegt]{Peter Bouwknegt}

\address[Peter Bouwknegt]{
Department of Physics and Mathematical Physics, 
and Department of Pure Mathematics \\
University of Adelaide \\
Adelaide, SA 5005 \\
Australia}
\email{pbouwkne@physics.adelaide.edu.au, 
       pbouwkne@maths.adelaide.edu.au}

\author[KC Hannabuss]{Keith Hannabuss}

\address[Keith Hannabuss]{
Mathematical Institute\\
University of Oxford\\
England}
\email{kch@balliol.oxford.ac.uk}

\author[V Mathai]{Varghese Mathai}

\address[Varghese Mathai]{
Department of Pure Mathematics \\
University of Adelaide \\
Adelaide, SA 5005 \\
Australia}
\email{vmathai@maths.adelaide.edu.au}


\begin{abstract}
In this paper, we initiate the study of
$C^*$-algebras $\alg$ endowed with a twisted action of a locally
compact Abelian Lie group $\sfG$, and we construct a  twisted crossed product $\alg\rtimes \sfG$,
which is in general a nonassociative, noncommutative, algebra.
The duality properties of this twisted
crossed product algebra are studied in detail, and are applied to T-duality in Type II string theory
to obtain the T-dual of a general principal torus bundle with general  H-flux, which we will
argue to be a bundle of noncommutative, nonassociative tori. We also show that this
construction of the T-dual includes all of the special cases that were previously analysed.
\end{abstract}
\maketitle

\section{Introduction}

Recent work has revealed the strong connections between T-duality in
string theory and Takai duality for C*-algebras, but for general
H-fields $C^*$-algebras are no longer adequate. In this paper we present a
generalisation which permits a very precise description of the
general T-dual.

Let $\sfT$ be a compact connected Abelian Lie group of rank $\ell$ with Lie algebra $\ft$, 
and $E \to M$ a principal $\sfT$-bundle with connection. 
By the Chern-Weil construction the space $\Omega(E)^\sfT$ of $\sfT$-invariant forms on
$E$ is isomorphic to the space of forms on $M$ with values in $\wedge \widehat{\ft}$, i.e.
\begin{equation} \label{eqAa}
\Omega^k(E)^\sfT \ \cong\ \bigoplus_{p=0}^k \Omega^p(M,\wedge^{k-p}\ \widehat{\ft})\,,
\end{equation}
and by a classical result of Chevalley and Koszul,
the de-Rham complex $(\Omega^\bullet(E),d)$ is chain homotopy equivalent to the 
complex $(\Omega^\bullet(M, \wedge^\bullet \widehat \ft), D)$ with a modified de-Rham
differential $D$, and hence that the associated cohomologies are isomorphic. 
Furthermore, we know that any class in $H^k(E)$ can be represented
by a form in $\Omega^k(E)^\sfT$.
Thus, under this isomorphism, an $H$-field in $H^3(E)$ can be considered as a 4-tuple
$H=(H_3,H_2,H_1,H_0)$ with
$H_p\in \Omega^p(M,\wedge^{3-p}\ \widehat{\ft})$ for $p=0, 1, 2, 3$, closed under the 
action of $D$ (cf. \cite{GHV,BHMb} for more details).
[The conclusions in 
this paper are valid for integral classes $H\in H^3(E,\ZZ)$ as well, since this paper
deals exclusively with the introduction of the additional `degree of freedom'  $H_0$,
which does not carry torsion, on top of established results which hold in the case of 
torsion $H$.  
Note that $H_0$ can be identified with the restriction of $H$ to a fibre.
For simplicity we have chosen to 
formulate some of the results in terms of differential forms.]

T-duality for principal circle bundles  was treated geometrically in \cite{BEMa, BEMb}, 
and dimensional considerations force the $H_0$ and  $H_1$
components of the H-flux to vanish in this case. The T-dual turns out to be another principal 
circle bundle with T-dual H-flux.
The arguments were extended in \cite{BHMa} to principal $\torus^\ell$-bundles with H-flux satisfying 
condition that the $H_0$ and  $H_1$ components vanish. Then the T-dual turns out to be another
principal $\torus^\ell$-bundle with T-dual H-flux having vanishing $H_0$ and  $H_1$ components.
The analysis in \cite{MR, MRb} shows that if one considers principal $\torus^\ell$-bundles with H-flux satisfying condition that just the $H_0$ component vanishes,
then one arrives at the surprising conclusion that the T-dual bundle has to 
have noncommutative tori as fibres, provided the $H_1$ component 
is non-zero. The weaker condition in \cite{MR, MRb} permits non-vanishing $H_1$, but still
excludes non-zero $H_0$.
In this paper we shall remove the last of these constraints, to allow a
non-vanishing $H_0$ component.
In this case, we arrive at the astonishing conclusion that 
the T-dual bundle has to have {\em nonassociative} tori as fibres, taking it even
beyond the normal range of non-commutative geometry.

The key step is provided by a new explicit construction of a continuous
trace algebra $\cb$ having a given Dixmier-Douady invariant, whose spectrum is
the total space $E$ of the principal $\sfT$-bundle, together with automorphisms
$\beta_g$ for $g\in \sfG=\ft$, which transform the spectrum in a way compatible
with the $\sfG$-action on $E$.
The new features arise because in general $g \mapsto \beta_g$ is not a
homomorphism but satisfies $\beta_x\beta_y = \ad(v(x,y))\beta_{xy}$, where
$\ad(v)$ denotes conjugation by a unitary element of the multiplier
algebra $M\cb$.
One expects the algebra associated with the T-dual to be the crossed
product $\cb\rtimes_{\beta}\sfG$, but the twisting forces us to take a
suitably twisted crossed product $\cb\rtimes_{\beta,v}\sfG$.
Such (Leptin-)Busby-Smith twistings have long been known but, for
non-trivial $H_0$,  $v(x,y)$ is not  a cocycle and that means that associativity
fails in the twisted crossed product.

In Section 2 we give the relationship between the differential forms and
the multicharacters on $\sfG$ which will be used in our later constructions.
Section 3 reviews the generalised Busby-Smith twisted crossed products.
An example of a twisting is given in Section 4, together with a proof that
it is the only type up to stability.
This is followed by an example of a non-associative generalisation of the
compact operators.

The theory of twisted induced representations is developed in Section 6 and
then used to construct examples of algebras with given spectrum and
Dixmier--Douady class in Section 7.
In Section 8 it is shown that the twisted crossed product of
a twisted induced algebra is isomorphic to a generalisation of the twisted
compact operators.
In Section 9, the double dual is shown to be the tensor product of the original
algebra with the twisted compact operators, that is, Morita equivalent in 
this category to the original algebra. In Section 10, the mathematical results 
of the previous sections are used to justify the assertion that 
the T-dual to a general principal torus bundle with H-flux, is a bundle of nonassociative tori. 
The final section outlines how associativity can be restored by working
in a different category, an idea which will be explored in more detail
in a subsequent paper.


\section{Differential forms and multicharacters}

Let $\sfT$ be a compact connected Abelian Lie group of rank $\ell$, 
and $E \to M$ a principal $\sfT$-bundle.
In essence the action of $\sfT$  on $E$ provides a map from the Lie algebra 
$\ft$ to vector fields on $E$, which we write $X \mapsto \xi_X$, and then  a 
$p$-form $f \in \Omega^p(E)$ in the fibre directions defines the antisymmetric 
multilinear form valued function on $M$
\begin{equation}
(\xi^*f)(X_1,X_2,\ldots,X_p) =
f(\xi_{X_1},\xi_{X_2},\ldots,\xi_{X_p}).
\end{equation}
For abelian Lie groups the form exponentiates to a multicharacter  on 
$\sfG=\ft$
\begin{equation}
\phi \left( \exp(X_1),\ldots,\exp(X_p) \right)
= \exp(-2\pi i(\xi^*f)(X_1,\ldots,X_p)).
\end{equation}
(A multicharacter is a character in each variable, and this property 
follows from the additivity of $\xi^*f$, and $\phi$ is also antisymmetric
in the sense that even permutations of its variables leave it unchanged
whilst odd permutations invert it.)

The multicharacter property ensures that this is always a (Moore) cocycle
in $Z^p(\ft,\torus)$, since, for example when $p=3$,
\begin{align*}\phi(y,z,w)\phi(x,yz,w)\phi(x,y,z)
&= \phi(y,z,w)\phi(x,y,w)\phi(x,z,w)\phi(x,y,z)\\
&= \phi(xy,z,w)\phi(x,y,zw).\end{align*}
(It is known that every cohomology class in $H^2(\real^n,\torus)$ can be 
represented by an antisymmetric bicharacter of the form $\phi$ \cite{Kle, Ha}, 
and for $p=3$ it is certainly true that smooth cocycles are cohomologous to 
antisymmetric tricharacters.)

\begin{example}
Consider the torus bundle $\torus^3$ over a point, with $H_0$ the class 
defined by $k$ times the volume form $dx^1\wedge dx^2 \wedge dx^3$.
The associated antisymmetric form on ${\bf a},{\bf b},{\bf c} \in
\ft = \real^3$ is then given by 
\begin{equation}
f({\bf a},{\bf b},{\bf c}) = k[{\bf a},{\bf b},{\bf c}] 
\equiv k{\bf a.}({\bf b}\times{\bf c}).
\end{equation}
whence 
$\phi({\bf a},{\bf b},{\bf c}) = \exp(-2k\pi i[{\bf a},{\bf b},{\bf c}])$.
\end{example}

Although we are mainly interested in the abelian groups $\sfT = \torus^n$, 
$\sfG= \ft = \real^n$, and $\sfN \cong \integer^n$ the kernel of the 
exponential map $\ft \to \sfT$, the constructions which we present in 
Chapters 3 to 8 are 
valid for general unimodular separable locally compact groups with a 
tricharacter $\phi$.
For that reason we shall write the group composition multiplicatively, and be 
careful not to commute terms.
However, this generalisation is less sweeping than may appear because the 
tricharacter $\phi$ on $\sfG$ defines a homomorphism of each variable into the 
abelian group $\torus$, and so must be lifted from a tricharacter on the 
abelianisation $\sfG/[\sfG,\sfG]$. 

Finally we note that, by definition, the Dixmier-Douady class has components which 
are integral 3-forms, and that means, in particular, that the tricharacter 
$\phi$ constructed from $H_0$ is identically 1 on 
$\sfN\times \sfN\times \sfN$, where 
$\sfN$ is the kernel of the exponential map. 
We shall assume this to be true in the general case.


\section{Generalised Busby-Smith twisted crossed products}

As noted above we now work with a general unimodular separable locally compact 
groups $\sfG$ and a closed subgroup $\sfN$ on which the tricharacter $\phi$ 
has trivial restriction.
For any group $\sfG$ one can interpret $H^2(\sfG,\sfA)$ as classifying central 
extensions of $\sfA$ by $\sfG$, whilst, for $p>3$, $H^p(\sfG, A)$ is usually 
interpreted in terms of crossed modules, 
\cite{EM1, EM2, Bro, Ho, Hu, Rat, McL}, with elements of 
$H^3(\sfG,\sfA)$ known as MacLane--Whitehead obstructions, \cite{Whi, MW}.
However, we shall see that these classes also arise in a $C^*$-algebraic 
context.

The $H$ field is usually linked to the equivariant Brauer group of a 
continuous trace C$^*$-algebra $\alg$ with spectrum $E$, on which a group 
$\sfG$ acts as automorphisms so that the dual action $E$ agrees with the 
bundle structure.
However, the equivariant Brauer group can be described entirely in terms of 
cohomology classes in $H^p(M,H^{3-p}(\sfG,\torus))$ for $p= 1,2,3$, 
\cite{CKRW}, leaving 
no room for $H$-fields with a component in $H^0(M, H^3(\sfG,\torus))$, (for 
example, any non-trivial $H$-field on $\sfG$ considered as a principal 
$\sfG$-bundle over a point).
Since the representatives of $H^0$ are locally constant functions we shall
concentrate our attention on $H^3(\sfG,\torus)$.

This suggests that we must consider a wider class of algebras or actions.
In fact, inner automorphisms autmatically act trivially on the spectrum, so 
that only homomorphisms of $\sfG$ to the outer automorphisms $\out(\alg) = 
\aut(\alg)/\inn(\alg)$ are interesting.
However, to work with these one needs a lifting $\alpha: \sfG \to \aut(\alg)$.
The problem then is that $\alpha_x\alpha_y$ and $\alpha_{xy}$ can differ by an 
inner automorphism $\ad(u(x,y)): a \mapsto u(x,y)au(x,y)^{-1}$, that is 
$\alpha_x\alpha_y = \ad(u(x,y))\alpha_{xy}$.
We can take $u(x,y) = 1$ whenever $x$ or $y$ is the identity.

This is almost precisely the data needed to define a Busby--Smith 
(or Leptin) twisted crossed product $\alg\rtimes_{\alpha,u}\sfG$ of $\alg$ and 
$\sfG$, \cite{Lep, BS, RSW}.
Assuming that $u$ is a measurable function on $\sfG\times \sfG$, we can define 
a twisted convolution product and adjoint on $C_0(\sfG,\alg)$ by
\begin{align*}(f*g)(x) &= \int_G f(y)\alpha_y[g(y^{-1}x)]u(y,y^{-1}x)\,dy,\\
f^*(x) &= u(x,x^{-1})^{-1}\alpha_x[f(x^{-1})]^*\end{align*}
and complete this to get a new algebra.

The link with the algebraists' picture of $H^3(\sfG,\sfA)$ arises because $u$ is 
no longer a cocycle since the condition linking $\alpha$ and $u$ tells us only 
that the adjoint actions of $u(x,y)u(xy,z)$ and $\alpha(x)[u(y,z)]u(x,yz)$ 
coincide, so that one has a modified cocycle condition: 
\begin{equation}
\phi(x,y,z)u(x,y)u(xy,z) = \alpha_x[u(y,z)]u(x,yz)
\end{equation}
for some central unitary element $\phi(x,y,z)\in UZ(\alg)$.
It is easy to check that $\phi$ is a cocycle defining an element of
$H^3(\sfG, UZ(\alg))$.
(Essentially the same argument is used in \cite{Car} to explain the origin of 
the Gauss anomaly and Jackiw's non-associative anomaly in quantum field
theory.)
When $\phi$ is a tricharacter one can still form a twisted crossed product.

\begin{proposition}
When $\phi$ defined as above is an antisymmetric tricharacter the twisted 
crossed product $\alg\rtimes_{\alpha,u}\sfG$ satisfies the $*$-algebra 
identity 
$(f*g)^* = g^{*}*f^*$, and is associative if and only if $\phi \equiv 1$.
In fact, we have
\begin{align*}((f*g)*h)(x)
&= \int_{\sfG\times \sfG} f(z)\alpha_z[g(z^{-1}y)]\alpha_y[h(y^{-1}x)]
u(z,z^{-1}y)u(y,y^{-1}x)\,dydz,\\
(f*(g*h))(x)
&= \int_{\sfG\times \sfG} f(z)\alpha_z[g(z^{-1}y)]\alpha_{y}[h(y^{-1}x)]
\phi(z,z^{-1}y,y^{-1}x)
u(z,z^{-1}y)u(y,y^{-1}x)\,dydz.\end{align*}
\end{proposition}
 
\begin{proof}
Using the modified cocycle identity, one calculates that
\begin{equation}
(f*g)^*(x) = \int_\sfG\phi(x,x^{-1}y,y^{-1})^*u(y,y^{-1})^*
\alpha_y[g(y^{-1})]^*u(x,x^{-1}y)^*\alpha_x[f(x^{-1}y)]^*\,dy,
\end{equation}
whilst
\begin{equation}
(g^**f^*)(x) = \int_G\phi(y,y^{-1}x,x^{-1}y)^*u(y,y^{-1})^*
\alpha_y[g(y^{-1})]^*u(x,x^{-1}y)^*\alpha_x[f(x^{-1}y)]^*\,dy,
\end{equation}
and for antisymmetric tricharacters both factors invoving $\phi$ are 1.

The twisted crossed product algebra $\alg\rtimes_{\alpha,u} \sfG$ has
\begin{align*}
((f*g)*h)(x)
&= \int_\sfG (f*g)(y)\alpha_y[h(y^{-1}x)]u(y,y^{-1}x)\,dy,\\
&= \int_{\sfG\times \sfG} f(z)\alpha_z[g(z^{-1}y)]u(z,z^{-1}y)
\alpha_y[h(y^{-1}x)]u(y,y^{-1}x)\,dydz\\
&= \int_{\sfG\times \sfG} f(z)\alpha_z[g(z^{-1}y)]
\alpha_z\alpha_{z^{-1}y}[h(y^{-1}x)]u(z,z^{-1}y)u(y,y^{-1}x)\,dydz\\
&= \int_{\sfG\times \sfG} f(z)\alpha_z[g(z^{-1}y)]\alpha_{z^{-1}y}[h(y^{-1}x)]
u(z,z^{-1}y)u(y,y^{-1}x)\,dydz,\end{align*}
and, using the modified cocycle identity,
\begin{align*}(f*(g*h))&(x)
= \int_\sfG f(z)\alpha_z[(g*h)(z^{-1}x)]u(z,z^{-1}x)\,dz\\
&= \int_{\sfG\times \sfG}
f(z)\alpha_z[g(z^{-1}y)]\alpha_{z^{-1}y}[h(y^{-1}x)]u(z^{-1}y,y^{-1}x)
u(z,z^{-1}y)\,dydz,\\
&= \int_{\sfG\times \sfG}
f(z)\alpha_z[g(z^{-1}y)]\alpha_{z^{-1}y}[h(y^{-1}x)]
\alpha_z[u(z^{-1}y,y^{-1}x)]u(z,z^{-1}y)\,dydz\\
&= \int_{\sfG\times \sfG}
f(z)\alpha_z[g(z^{-1}y)]\alpha_{z^{-1}y}[h(y^{-1}x)]\phi(z,z^{-1}y,y^{-1}x)
u(z,z^{-1}y)u(y,y^{-1}x)\,dydz,
\end{align*}
so that the Busby-Smith twisted crossed product is non-associative except in 
the case $\phi \equiv 1$.
\end{proof}

Henceforth we shall always take $\phi$ to be an antisymmetric tricharacter.

Usually Busby-Smith products are only defined when $\phi=1$, but we shall see 
that much of the theory goes through without that assumption, so that this 
provides a means of constructing non-associative from associative algebras.
The non-associativity becomes even more transparent when one considers a 
covariant representation $(U,\pi)$ of $(\sfG,\alg)$ satisfying the conditions
\begin{equation}
U(x)\pi(a)U(x)^{-1} = \pi(\alpha_x(a)), \qquad U(x)U(y) =
\pi(u(x,y))U(xy).
\end{equation}
These give
\begin{align*}U(x)[U(y)U(z)]  &=  U(x)[\pi(u(y,z))]U(yz)\\
&= \pi(u(x,yz))\pi(\alpha_xu(y,z))U(x(yz))\end{align*}
\begin{align*}[U(x)U(y)]U(z) &= \pi(u(x,y))U(xy)U(z)\\
&= \pi(u(x,y))\pi(u(xy,z))U((xy)z),\end{align*}
so that 
\begin{equation}
\phi(x,y,z)U(x)[U(y)U(z)]  = [U(x)U(y)]U(z).
\end{equation}
In fact, $H^3(\sfG)$ was already interpreted as defining a nonassociative 
structure in \cite{EM2}, and this has resurfaced in the physics literature 
\cite{Jac, CS}.


\section{Generalised Packer--Raeburn stabilisation}

For a given antisymmetric tricharacter $\phi$ on $\sfG$ there is 
a simple example of an algebra with twisting on which 
$\sfG$ acts as automorphisms.
It is derived from the imprimitivity algebra generated by multiplication and 
translation operators on $L^2(\sfG)$.
The right regular representation $\rho$ acts on $\psi\in L^2(\sfG)$ by
$(\rho(x)\psi)(v) = \psi(vx)$, and  we define
\begin{equation}
(u_\rho(y,z)\psi)(v) = \phi(v,y,z)\psi(v).
\end{equation}
(We shall often be interested in the case when $\sfG$ is a closed subgroup of 
a group $\sfH$, with $\phi$ defined on $\sfH\times \sfH\times \sfH$ and 
trivial on $\sfG\times \sfG\times \sfG$.
Then $u_\rho(y,z)$ can be defined as a multiplication operator on $L^2(\sfG)$ 
for general $y$, $z\in \sfH$, and the restriction of $u_\rho$ to 
$\sfG\times \sfG$ is identically 1.)

Now
\begin{align*}
\phi(x,y,z)(u_\rho(x,y)u_\rho(xy,z)u_\rho(x,yz)^{-1}\psi)(v)
&= \phi(x,y,z)\phi(v,x,y)\phi(v,xy,z)\phi(v,x,yz)^{-1}\psi(v)\\
&= \phi(vx,y,z)\psi(v)
\end{align*}
and
\begin{equation}
(\rho(x)u_\rho(y,z)\rho(x)^{-1}\psi)(v) = (u_\rho(y,z)\rho(x)^{-1}\psi)(vx)
= \phi(vx,y,z)(\rho(x)^{-1}\psi)(vx) = \phi(vx,y,z)\psi(v),
\end{equation}
so that, setting $\alpha_x = \ad(\rho(x))$, we get
$\phi(x,y,z)u_\rho(x,y)u_\rho(xy,z)u_\rho(x,yz)^{-1} = \alpha_x[u_\rho(y,z)]$.
This gives an explicit realisation of an algebra $\cc = C_0(\sfG)$ with an
action of $\sfG$ by automorphisms and with the appropriate Busby-Smith
obstruction.
(When $\sfG$ is non-compact $u_\rho(x,y)$ is in the multiplier algebra rather 
than the algebra itself.)
The same idea can be extended to the right regular $\sigma$-representation
on $L^2(\sfG)$ given by
$(\rho(x)\psi)(v) = \sigma(v,x)\psi(vx)$ for any borel multiplier $\sigma$,
and this makes no difference to the cocycle identity.

Naturally, one can also work with the left regular representation 
$(\lambda(x)\psi)(v) = \psi(x^{-1}v)$, and 
\begin{equation} \label{eqqDc}
(u_\lambda(y,z)\psi)(v) = \phi(v,y,z)^{-1}\psi(v).
\end{equation}
This is useful because it links directly to the formulation used by \cite{PR} 
to show that twisted crossed products defined by cocycles are stably 
equivalent to normal crossed products.
In our case $u$ is not a cocycle, but there is nonetheless a nice 
generalisation of the Packer--Raeburn Theorem.

\begin{theorem} \label{thDa}
Let $\alg$, $\sfG$, $\alpha$, $u$ be as above.
There exists a strongly continuous action $\beta$ of $\sfG$ on 
$\alg\otimes\cpt(L^2(\sfG))$ and a twisting $u_\lambda$ such that 
$(\beta,u_\lambda)$ are exterior 
equivalent to $(\alpha\otimes{\rm id}, u\otimes 1)$, that is, there exists 
$v_s = (1\otimes \lambda_s)({\rm id}\otimes M(u(s,\cdot))^*)$ such that
\begin{equation}
\beta_s = \ad(v_s)(\alpha_s\otimes{\rm id}), \qquad
{\rm id}\otimes u_\lambda(s,t) = v_s\alpha_s(v_t)(u(s,t)\otimes 1)v_{st}^*.
\end{equation}
\end{theorem}

\begin{proof}
The proof of Theorem 3.4 in \cite{PR} is still valid as far as the last line 
on page 301.
At that point the original argument uses the fact that $u$ is a cocycle to 
show that it has been untwisted by the exterior equivalence.
In our case $u$ satisfies a modified cocycle identity, so that that last line 
(in the original notation) gives $\phi(s,t,r)^{-1}\xi(r) = 
(u_\lambda(s,t)\xi)(r)$.
\end{proof}

One similarly obtains:
\begin{corollary}\label{cor: stab}
In the situation of Theorem 4.1 one has
\begin{equation}
(\alg\rtimes_{\alpha,u}\sfG)\otimes\cpt(L^2(\sfG)) \cong
(\alg\otimes\cpt)\rtimes_{\beta,u_\lambda}\sfG \,.
\end{equation}
\end{corollary}

The Packer-Raeburn stabilisation trick is used to show that up to 
stabilisation by tensoring with compact operators 
$\alg\rtimes_{\alpha,u}\sfG$ and $\alg\rtimes_{\beta,u_\lambda}\sfG$ are 
isomorphic.
The original idea derived from Quigg's generalisation of Takai duality for 
twisted crossed products, and their $\beta$ is just equivalent to Quigg's 
double dual $\widehat{\widehat{\alpha}}$. 
One may therefore obtain a duality theorem by the same procedure.
We postpone discussion of this until Section 9 where we shall give a much more 
detailed account of duality for abelian groups.

\section{Twisted compact operators}

Before developing the theory further it is useful to give a very simple 
example of a non-associative algebra, obtained by twisting the algebra of 
compact operators on $L^2(\sfG)$ using the factor $\phi(x,y,z)$.
We start with the Hilbert-Schmidt operators, realised as kernels $K(x,y)$ for 
$x,y\in \sfG$, with the involution $K^*(x,y) = \conj{K(y,x)}$, and norm
\begin{equation}
\|K\|_{HS}^2 = \int_{\sfG\times \sfG} |K(x,y)|^2
\end{equation}
and define the new multiplication
\begin{equation}
(K_1*K_2)(x,z) = \int_{y\in \sfG} \phi(x,y,z)K_1(x,y)K_2(y,z)\,dy.
\end{equation}
This is consistent with the involution because
\begin{align*}(K_1*K_2)^*(x,z) &= \conj{(K_1*K_2)(z,x)}\\
&= \int_{\sfG} \phi(z,y,x)^{-1}\conj{K_1(z,y)}\,\conj{K_2(y,x)}\,dy\\
&= \int_{\sfG} \phi(x,y,z)K_2^*(x,y)K_1^*(y,z)\,dy\\
&= (K_2^**K_1^*)(x,z).\end{align*} 
The fact that $\phi(x,y,x) = 1$ also means that we still have
\begin{equation}
\|K\|_{HS}^2 = \int_\sfG (K^**K)(x,x)\,dx.
\end{equation}
As usual, one can define a $C^*$-norm using the left regular representation
$\|K_1\| = \sup(\|K_1*K_2\|_{HS}/\|K_2\|_{HS})$.
By using the Cauchy-Schwarz inequality and by considering rank one projections one 
sees that this is equivalent to the ordinary operator norm.
The twisted compact operators $\cpt_\phi(L^2(\sfG))$ are the completion of the 
Hilbert-Schmidt operators with respect to that norm.

Unfortunately the new multiplication is not associative unless
\begin{equation}
\phi(x,y,z)\phi(x,z,w) = \phi(x,y,w)\phi(y,z,w),
\end{equation}
for all $x,y,z,w \in \sfG$.
In that case $\phi(x,y,z) = \phi(x,y,w)\phi(y,z,w)/\phi(x,z,w)$, but 
conversely, whenever $\phi(x,y,z)$ has the form $\psi(x,y)\psi(y,z)/\psi(x,z)$,
 for some function $\psi$, the algebra is associative.
The twisted multiplication of kernels was used in \cite{CHMM} to study the 
hyperbolic quantum Hall effect, but in that two-dimensional situation the 
multiplication is automatically associative. Once one gets to three-dimensions 
the analogous algebra is non-associative.

\begin{proposition}
The group $\sfG$ acts on the twisted algebra $\cpt_\phi(L^2(\sfG))$ 
with multiplication 
\begin{equation}
(K_1*K_2)(x,z) = \int_\sfG \phi(x,y,z)K_1(x,y)K_2(y,z)\,dy.
\end{equation}
by natural $*$-automorphisms
\begin{equation}
\theta_x[K](z,w) = \phi(x,z,w)K(zx,wx),
\end{equation}
and $\theta_x\theta_y = \ad(\sigma(x,y))\theta_{xy}$
where $\ad(\sigma(x,y))[K](z,w) = \phi(x,y,z)\phi(x,y,w)^{-1}K(z,w)$ comes from 
the multiplier $\sigma(x,y)(v) = \phi(x,y,v)$.
\end{proposition}

\begin{proof}
Using the tricharacter property of $\phi$, we see that
\begin{align*}(\theta_x[K_1]*\theta_x[K_2])(z,w)
&= \int_\sfG \phi(z,v,w)\phi(x,z,v)\phi(x,v,w)
K_1(zx,vx)K_2(vx,wx)\,dv\\
&= \int_\sfG \phi(zx,vx,wx)
\phi(x,z,w)
K_1(zx,vx)K_2(vx,wx)\,dv\\
&= \phi(x,z,w)
(K_1*K_2)(zx,wx),\\
&= \theta_x[K_1*K_2](z,w).\end{align*}
We also have
\begin{equation}
\theta_x[K]^*(z,w) = \conj{\theta_x[K](w,z)}
= \conj{\phi(x,z,w)K(w,z)}
= \phi(x,w,z)K^*(w,z)
= \theta_x[K^*](z,w).
\end{equation} 
Moreover,
\begin{align*}\theta_x\theta_y[K](z,w) &=
\phi(x,z,w)(\theta_y[K])(zx,wx)\\
&= \phi(x,z,w)\phi(y,zx,wx)K(zxy,wxy)\\
&= \phi(x,y,z)\phi(x,y,w)^{-1}\theta_{xy}[K](z,w)\\
&= \ad(\sigma(x,y))[\theta_{xy}[K]](z,w),\end{align*}
with $\sigma(x,y)(z,w) = \phi(x,y,z)\delta(z-w)$.
\end{proof}

\noindent
{\it Note:}
There is also a left-handed version of this which uses the multiplication
\begin{equation}
(K_1*K_2)(x,z) = \int_{\sfG} \phi(x,y,z)^{-1}K_1(x,y)K_2(y,z)\,dy.
\end{equation}
and automorphisms
\begin{equation}
\tau_x[K](z,w) = \phi(x,z,w)K(x^{-1}z,x^{-1}w),
\end{equation}
and it is this version which will appear later, in Section 9.

\bigskip
When $\sfG$ is a contractible group the twisted compact operators are just a 
deformation of the usual ones.

\begin{proposition}
When $\sfG$ is a contractible group $\cpt_\phi(L^2(\sfG))$ is a continuous 
deformation of $\cpt(L^2(\sfG))$.
\end{proposition}

\begin{proof}
Let $\{\epsilon_t: t\in [0,1]\}$ give a contraction of $G$ onto the identity, 
that is $\epsilon_t:\sfG \to \sfG$ is continuous and satisfies 
$\epsilon_0(x) = x$ and $\epsilon_1(x)$ is the identity for all $x\in G$.
We then define (for the right-handed version)
\begin{equation}
(K_1*_tK_2)(x,z) 
= \int_{\sfG} \phi(x,\epsilon_t(y),z)K_1(x,y)K_2(y,z)\,dy,
\end{equation}
so that at $t=0$ we have the twisted and at $t=1$ the untwisted product.
\end{proof}

\section{Twisted induced algebras}

The introduction of $u$ and $\phi$ has far-reaching consequences, 
because almost all the standard procedures have to be deformed, and we shall 
now investigate these in more detail.

Suppose that $\alg$ is a C$^*$-algebra on which $C(M)$ acts as double centralisers and  
the subgroup $\sfN$ of $\sfG$ acts by 
automorphisms $\alpha_r$, ($r\in \sfN$), and that for each $s,t\in \sfN$ there 
are unitaries $u(s,t)$ in the multiplier algebra $M(\alg)$ such that 
$\alpha_s\alpha_t =\ad(u(s,t))\alpha_{st}$, and also the modified cocycle 
condition (for a specified continuous tricharacter $\phi$)
\begin{equation}
\alpha_r[u(s,t)] u(r,st) = \phi(r,s,t)u(r,s)u(rs,t).
\end{equation}
Such algebras always exist, since we can take the algebra freely generated by 
a collection of symbols $\{u(s,t): s,t\in \sfN\}$ and define the automorphism 
$\alpha_r$ by the formula
\begin{equation}
\alpha_r[u(s,t)]  = \phi(r,s,t)u(r,s)u(rs,t)u(r,st)^{-1}.
\end{equation}

We shall suppose that $u$ and $\phi$ extend to continuous functions on $\sfG$ 
satisfying the same relations:
\begin{equation}
\alpha_r[u(x,y)] u(r,xy) = \phi(r,x,y)u(r,x)u(rx,y),
\end{equation}
for $r\in \sfN$ and $x, y\in \sfG$.
(We are mainly interested in the case when $\sfN$ is a maximal rank lattice in 
a vector group $\sfG$, and then $\phi$ automatically extends, and in the 
interesting examples $u$ does too.)
Normally one would induce an algebra admitting a $\sfG$-action from that
containing an $\sfN$-action, but that will no longer work since the induced 
algebra is trivial. 
Instead we consider the $u$-induced algebra $\cb=\uind_\sfN^\sfG\alg$ 
described in the next result.

\begin{proposition}
The space $\cb=\uind_\sfN^\sfG\alg$ of functions $f \in C_0(\sfG,\alg)$ which 
satisfy 
$f(rx) = \ad(u(r,x))^{-1}\alpha_r[f(x)]$ for all $x\in \sfG$ and $r\in \sfN$ 
is not trivial, and is closed under pointwise multiplication of functions 
$(f_1f_2)(x) = f_1(x)f_2(x)$ and the involution $f^*(x) = f(x)^*$.
The norm $\|f\| = \sup\|f(x)\|$ is a C$^*$-norm.
\end{proposition}

\begin{proof}
We first note that (using the cocycle condition and remembering that the
adjoint action is unaffected by the central factor $\phi$) functions in
the space satisfy
\begin{align*}f(rsx) &= \ad(u(r,sx))^{-1}\alpha_r[f(sx)] \\
&= \ad(u(r,sx))^{-1}\alpha_r[\ad(u(s,x))]^{-1}\alpha_s[f(x)] \\
&= \ad(u(r,sx))^{-1}\alpha_r[\ad(u(s,x))]^{-1}\alpha_r\alpha_s[f(x)] \\
&= \ad(u(rs,x))^{-1}\ad(u(r,s))^{-1}\alpha_r\alpha_s[f(x)] \\
&= \ad(u(rs,x))^{-1}\alpha_{rs}[f(x)] ,\end{align*}
showing consistency of the condition.
Without $u$ this consistency check would fail and the induced algebra would 
be trivial, showing why one cannot use the normal induced algebra.
We shall exhibit some useful explicit functions in the induced algebra in 
Proposition 6.4, but there is also a general construction which is useful.
For $f$ a function $C_0(\sfG)$ and $a$ an element in $\alg$, we define an 
$\alg$-valued function $(f\diamondsuit a)$ on $\sfG$ by
\begin{equation}
(f\diamondsuit a)(x) = \int_\sfN f(nx)\alpha_n^{-1}[\ad(u(n,x))[a]]\,dn.
\end{equation}
Using the cocycle identity for $\ad(u)$ (the obstruction $\phi$ is central 
and so disappears in the adjoint action) we then check that 
\begin{align*}(f\diamondsuit a)(rx) 
&= \int_\sfN f(nrx)\alpha_n^{-1}[\ad(u(n,rx))[a]]\,dn\\
&= \int_\sfN f(nrx)\alpha_n^{-1}[\ad(\alpha_n[u(r,x)])^{-1}\ad(u(n,r))
\ad(u(nr,x))[a]]\,dn\\
&= \ad(u(r,x))^{-1}\int_\sfN f(nrx)\alpha_n^{-1}[\ad(u(n,r))\ad(u(nr,x))[a]]\,dn\\
&= \ad(u(r,x))^{-1}\alpha_r\int_\sfN f(nrx)\alpha_{nr}^{-1}[\ad(u(nr,x))[a]]
\,dn\\
&= \ad(u(r,x))^{-1}\alpha_r[(f\diamondsuit a)(x)],\end{align*}
showing that $(f\diamondsuit a)$ defines an element of $\cb$.

Using the fact that $\alpha_r$ and $\ad(u(x,r))$ are automorphisms, we
see that
\begin{align*}(f_1f_2)(xr) = f_1(xr)f_2(xr)
&= \ad(u(x,r))\alpha_r[f_1(x)]\ad(u(x,r))\alpha_r[f_2(x)]\\
&= \ad(u(x,r))\alpha_r[f_1(x)f_2(x)]\\
&= \ad(u(x,r))\alpha_r[(f_1f_2)(x)]
\end{align*}
so that the space of $u$-induced functions is closed under the product.

Finally, exploiting the unitarity of $u(r,x)$, we have
\begin{align*}
f^*(rx) &= f(rx)^*\\
&= [u(r,x)^{-1}\alpha_r[f(x)]u(r,x)]^*\\
&= u(r,x)^{-1}\alpha_r[f(x)]^*u(r,x)\\
&= u(r,x)^{-1}\alpha_r[f^*(x)](r,x),\end{align*}
so that the involution respects the constraint.
Finally 
\begin{equation}
\|f^*f\| = \sup\|(f^*f)(x)\| 
= \sup\| f(x)^*f(x)\| 
= \sup\|f(x)\|^2
\end{equation} 
gives a $C^*$-norm.
\end{proof}

This shows that the induced space $\cb$ is actually a $*$-algebra (moreover,
an associative algebra,  since $\alg$ was associative).
A $C^*$-norm can be defined much as in the usual case.
Exploiting the ideas of the ordinary induced representations we can improve on 
this.

\begin{theorem}
If $\alg$ is a continuous trace algebra with spectrum $\widehat{\alg}$  then 
$\cb$ is a continuous trace algebra with spectrum 
$\widehat{\cb} = \sfN\backslash(\sfG\times 
\widehat{\alg})$, where the action of $r\in \sfN$  on 
$(x,\pi)\in \sfG\times\widehat{\alg}$ is defined by 
$r(x,\pi) = (rx,\pi\circ\alpha_{r}^{-1}\ad(u(r,x)))$.
\end{theorem}

\begin{proof}
We should start by checking that the above formula does indeed define an 
action, but that is essentially the same calculation just done to show that 
$f\diamondsuit a$ satisfies the equivariance condition for $\cb$.
The rest of the proof follows the ideas in \cite[6.16 to 6.21]{RW}, but with 
the actions on the other sides, and using our definition of $f\diamondsuit a$, 
and noting that in the proof of 6.18 one needs to restrict to neighbourhoods 
of both $s$ and $x$.
This enables us to show that the representation defined by $(x,\pi)\in 
\sfG\times\widehat{\alg}$, $(x,\pi): F\mapsto \pi(F(x)$ of $\cb$ is 
irreducible, 
because, for any $(x,a)\in \sfG\times \alg$ we can use appropriate 
$f\diamondsuit a$ to find $F\in \cb$ such that $F(x) = a$, and then use 
irreducibility of $\pi$.
The rest of the proof in \cite{RW} is independent of the twisting.
\end{proof}

We next introduce automorphisms of the twisted induced algebra.

\begin{proposition}
For $y\in \sfG$ and a function $f:\sfG \to \alg$ define 
$\beta_y[f](x) = \ad(u(x,y))[f(xy)]$.
Then $\beta_y$ preserves the subalgebra $\cb$ and defines a $*$-automorphism
of it.
\end{proposition}

\begin{proof}
\begin{align*}\beta_y[f](rx) &= \ad(u(rx,y))[f(rxy)]\\
&= \ad(u(rx,y))\ad(u(r,xy))^{-1}\alpha_r[f(xy)]\\
&= \ad(u(r,x))^{-1} \ad(\alpha_r[(u(x,y)]) \alpha_r[f(xy)]\\
&= \ad(u(r,x))^{-1}\alpha_r[\ad(u(x,y))] [f(xy)]\\
&= \ad(u(r,x))^{-1}\alpha_r[\beta_y[f](x)],\end{align*}
showing that $\beta_y$ satisfies the equivariance condition.

(To see that these are automorphisms we need only note that
\begin{align*}(\beta_y[f_1]\beta_y[f_2])(x)
&= (\beta_y[f_1](x)\beta_y[f_2])(x)\\
&= \ad(u(x,y))[f_1(xy)]\ad(u(x,y))[f_2(xy)]\\
&= \ad(u(x,y))[f_1(xy)f_2(xy)] \\
&= \ad(u(x,y))[(f_1f_2)(xy)] \\
&=\beta_y[f_1f_2](x),\end{align*}
as required, and compatibility with the involution is also easily checked.)
\end{proof}

Naturally the map $y\mapsto \beta_y$ is not a homomorphism.

\begin{proposition}
The functions $v(y,z):x \mapsto \phi(x,y,z)u(x,y)u(xy,z)u(x,yz)^{-1}$,
lie in the multiplier algebra of $\cb$ and satisfy
\begin{equation}
\beta_x[v(y,z)]v(x,yz) = \phi(x,y,z)v(x,y)v(xy,z).
\end{equation}
The automorphisms defined by $\beta$ satisfy the relations
\begin{equation}
\beta_y\beta_z = \ad(v(y,z))\beta_{yz}
\end{equation}
\end{proposition}

\begin{proof}
When $x\in \sfN$ we can write $v(y,z)(x) = \alpha_x[u(y,z)]$, but otherwise 
$\alpha_x$ is undefined, and we cannot reduce
$v(y,z)(x) =  \phi(x,y,z)^{-1}\ad(u(x,y))\ad(u(xy,z))\ad(u(x,yz))^{-1}$.
To check that $v(y,z)$ satisfies the equivariance condition for membership of 
$\cb$, we calculate
\begin{align*}u(rx,y)u(rxy,z)&u(rx,yz)^{-1}\\
&=
\phi(r,x,y)^{-1}u(r,x)^{-1}\alpha_r[u(x,y)]u(r,xy)u(rxy,z)u(rx,yz)^{-1}\\
&=
\phi(r,x,y)^{-1}\phi(r,xy,z)^{-1}u(r,x)^{-1}\alpha_r[u(x,y)]\alpha_r[u(xy,z)]u(r,xyz)u
(rx,yz)^{-1}\\
&=
\phi(r,x,y)^{-1}\phi(r,xy,z)^{-1}\phi(r,x,yz)u(r,x)^{-1}\alpha_r[u(x,y)u(xy,z)]\alpha_r[u(x,yz)]^{-
1}u(r,x)\\
&=
\phi(r,x,y)^{-1}\phi(r,xy,z)^{-1}\phi(r,x,yz)\ad(u(r,x))^{-1}\alpha_r[u(x,y)u(xy,z)u(x,yz)^{-1}]\\
&=
\phi(r,y,z)^{-1}\ad(u(r,x))^{-1}\alpha_r[u(x,y)u(xy,z)u(x,yz)^{-1}].\end{align*}
{}From this we see that
\begin{align*}[v(y,z)](rx) 
&= \phi(rx,y,z)\phi(r,y,z)^{-1}\ad(u(r,x))^{-1}
\alpha_r[\phi(x,y,z)^{-1}v(y,z)(x)]\\
&= \ad(u(r,x))^{-1}\alpha_r[v(y,z)(x)].\end{align*}
We cannot conclude that $v(y,z)$ lies in $\cb$ since it does not satisfy the 
analytic condition of vanishing outside compact sets, but it is certainly in 
the multiplier algebra.

By making some cancellations, we obtain
\begin{align*}\left(v(x,y)v(xy,z)v(x,yz)^{-1}\right)(s)
&= \frac{\phi(s,x,y)\phi(s,xy,z)}{\phi(s,x,yz)}
u(s,x)u(sx,y)u(sxy,z)u(s,xyz)^{-1}u(s,x)^{-1}\\
&= \phi(s,y,z)\ad(u(s,x))[u(sx,y)u(sxy,z)u(sx,yz)^{-1}]\\
&= \phi(x,y,z)^{-1}\ad (u(s,x))[v(y,z)(sx)]\\
&= \phi(x,y,z)^{-1}\beta_x[v(y,z)](s),\end{align*}
Finally we see that
\begin{align*}(\beta_y\beta_z[f])(x) &= \ad(u(x,y))[(\beta_z[f](xy)]\\
&= \ad(u(x,y))\ad(u(xy,z)[f(xyz)]\\
&= \ad(u(x,y))\ad(u(xy,z)\ad(u(x,yz)^{-1})[(\beta_{yz}f(x)]\\
&= \ad(v(y,z))(x)[(\beta_{yz}f(x)].\end{align*}
\end{proof}

\begin{corollary}
The action of $\sfG$ on $\widehat{\cb}$ defined by $\beta$ has orbit space
$\widehat{\cb}/\sfG = \widehat{\alg}/\sfN$, so that $\beta$ 
defines the principal $\sfG/\sfN$-bundle 
$(\sfG\times\widehat{\alg})/\sfN \to \widehat{\alg}/\sfN$.
\end{corollary}

\begin{proof}
The action of $y\in \sfG$ sends $\varpi\in \widehat{\cb}$ to 
$\varpi\circ\beta_y$.
In the earlier notation we have 
\begin{equation}
(x,\pi)(\beta_y[f]) = \pi(\beta_y[f](x))
= \pi(\ad(u(x,y))[f(xy)]),
\end{equation}
and, since inner automorphisms don't affect the class of a representation, 
this is equivalent to $(xy,\pi)$.
For $r\in \sfN$ this reduces to 
\begin{equation}
(x,\pi)(\beta_r[f]) = \pi(\ad(u(x,r)u(r,x)^{-1})\ad(u(r,x)\alpha_r[f(x)]),
\end{equation}
which is equivalent to $r(x,\pi)$, showing that the subgroup $\sfN$ stabilises 
the irreducible representations of $\cb$, so that we have a $\sfG/\sfN$ 
bundle, and that the orbit space is $\widehat{\alg}/\sfN$, as claimed.
\end{proof}


\section{Algebras with prescribed spectrum and Dixmier-Douady class}

Before tackling the general case it is useful to consider what happens for the 
principal bundle $\sfG/\sfN$ over a point.
In that case only the component $H_0$ can be non-trivial, and we assume that 
it defines the tricharacter $\phi$ as before. 

In fact \cite{CM} gives a universal construction for a principal projective unitary 
bundle over a group, but we shall give an alternative description of 
the algebra as a twisted induced algebra.
We shall induce from $\sfN$ to $\sfG$, and in order that the spectrum should 
be just 
$\sfG/\sfN$ we induce the algebra $\cpt(L^2(\sfN))$ of compact operators.
That carries the right regular representation $\rho$ of $\sfN$ and twisting
$u_\rho$ described in Section 4. 
Although the restriction of $u_\rho$ to $\sfN\times \sfN$ is 1, the extension 
to 
$\sfG\times \sfG$ gives the induced algebra a twist.

At this point it is instructive to consider why 
$\urind_\sfN^\sfG(\cpt(L^2(\sfN)))$ 
has non-vanishing Dixmier--Douady obstruction when the action of $\sfN$ on 
$\cpt(L^2(\sfN))$ is given by $\alpha_\rho = \ad\rho$. 
One might expect that the induced algebra $\urind_\sfN^\sfG(\cpt(L^2(\sfN))$ 
simply acts on the induced Hilbert space of square-integrable functions 
$\psi:\sfG \to L^2(\sfN)$, which satisfy the equivariance condition
\begin{equation}
\psi(rx) = u_\rho(r,x)^{-1}\rho(r)\psi(x),
\end{equation} 
so that there is no obstruction.
However, this is incorrect because the suggested equivariance condition on 
$\psi$ is inconsistent, when $u_\rho$ is not a cocycle:
\begin{align*}\psi(rsx) &= u_\rho(r,sx)^{-1}\rho(r)\psi(sx) \\
&= u_\rho(r,sx)^{-1}\rho(r)u_\rho(s,x)^{-1}\rho(s)\psi(x)\\
&= u_\rho(r,sx)^{-1}\alpha_r[u_\rho(s,x)]^{-1}\rho(r)\rho(s)\psi(x)\\
&= \phi(r,s,x)^{-1}u_\rho(rs,x)^{-1}u_\rho(r,s)^{-1}\rho(rs)\psi(x).\end{align*}
Now, as we noted in the last paragraph, $u_\rho(r,s) = 1$, but the presence of the 
argument $x\notin \sfN$ means that $\phi(r,s,x) \neq 1$, and we end up with 
constraints 
\begin{equation}
u_\rho(rs,x)^{-1}\rho(rs)\psi(x) = \psi(rsx) 
= \phi(r,s,x)^{-1}u_\rho(rs,x)^{-1}\rho(rs)\psi(x)
\end{equation}
which can be satisfied only by $\psi = 0$.

\begin{proposition}\label{thSevenone}
Let $\phi$ be the tricharacter of $\sfG$ constructed from $H_0$, $u_\rho$, 
$\rho$ defined on $L^2(\sfN)$ as in Section 4, and $\alpha_\rho = \ad\rho$.
The algebra $\urind_\sfN^\sfG(\cpt(L^2(\sfN)))$ has spectrum $\sfG/\sfN$, the 
$\sfG$ action on the spectrum is transitive with stabiliser $\sfN$, and the 
Dixmier-Douady class is described by the 3-form $H_0$.
\end{proposition}

\begin{proof}
We assume that $\phi$ is obtained from the class $f = H_0$ by the procedure 
described in Section 2.

Choose an open set $F \subseteq \sfG$ on which the projection $\pi$ to 
$\sfG/\sfN$ is 
one-one, and translates $F_i = F  x_i$ whose projections give a cover of 
$\sfG/\sfN$.
These translates share the property that the projection to $\sfG/\sfN$ is 
injective, 
and so we may choose sections $\gamma_i: \pi(F_i) \to \sfG$.
The differences $\gamma_{ij}(v) = \gamma_i(v)\gamma_j(v)^{-1}$ lie in $\sfN$.

The restriction of the induced algebra to algebra-valued functions on
$\pi(F_i) \subseteq \sfG/\sfN$ is Morita equivalent to $C(\pi(F_i))$ via the 
bimodule $X_i = L^2(\pi(F_i),L^2(\sfN))$, (restriction to the subsets enables 
us to sidestep the earlier problem with $L^2(\sfG,L^2(\sfN))$).
The actions are the obvious pointwise multiplicative actions, 
\begin{equation}
(f\psi)(v) = f(\gamma_i(v))\psi(v)
\end{equation}
and this is a imprimitivity bimodule in the sense of \cite{RW}.

Over the intersection $\pi(F_i)\cap\pi(F_j)$ there is an equivalence of the two 
bimodules $X_i$ and $X_j$, given by the map 
\begin{equation}
(g_{ij}\psi)(v) = u_\rho(\gamma_{ij}(v),\gamma_j(v))^{-1}
\rho(\gamma_{ij}(v))\psi(v)
\end{equation}
from $X_j$ to $X_i$.
To see why this works we note that
\begin{align*}(g_{ij}f\psi)(v)  
&= u_\rho(\gamma_{ij}(v),\gamma_j(v))^{-1}\rho(\gamma_{ij}(v))
f(\gamma_j(v))\psi(v)\\
&= \ad(u_\rho(\gamma_{ij}(v),\gamma_j(v))^{-1}
\ad(\rho(\gamma_{ij}(v))[f(\gamma_j(v))]\\
&\qquad u_\rho(\gamma_{ij}(v),\gamma_{j}(v))^{-1}\rho(\gamma_{ij}(v))\psi(v).\end{align*}
Since $\gamma_{ij}(v)\in \sfN$ this simplifies to 
\begin{equation}
f(\gamma_{ij}(v)  \gamma_j(v))
u_\rho(\gamma_{ij}(v),\gamma_j(v))^{-1}\rho(\gamma_{ij}(v))\psi(v) 
= (fg_{ij}\psi)(v).
\end{equation}

To compute the obstruction we must compare $g_{ij}g_{jk}$ and $g_{ik}$ on 
$\pi(F_i)\cap \pi(F_j)\cap \pi(F_k)$.
Now we have
\begin{align*}(g_{ij}g_{jk}\phi)(v)
&= u_\rho(\gamma_{ij}(v),\gamma_j(v))^{-1}\rho(\gamma_{ij}(v))(g_{jk}\psi)(v)\\
&= u_\rho(\gamma_{ij}(v),\gamma_j(v))^{-1}\rho(\gamma_{ij}(v))\\
&\qquad u_\rho(\gamma_{jk}(v),\gamma_k(v))^{-1}\rho(\gamma_{jk}(v))\psi(v)\\
&= u_\rho(\gamma_{ij}(v),\gamma_j(v))^{-1}\ad(\rho(\gamma_{ij}(v)))
[u_\rho(\gamma_{jk}(v),\gamma_k(v))^{-1}]\\
&\qquad \rho(\gamma_{ij}(v))\rho(\gamma_{jk}(v))\psi(v).\end{align*}
Applying the modified cocycle identity we have
\begin{align*}(g_{ij}g_{jk}\phi)(v)
&= \phi(\gamma_{ij}(v),\gamma_{jk}(v),\gamma_k(v))^{-1}
u_\rho(\gamma_{ik}(v),\gamma_k(v))^{-1}\\
&\qquad u_\rho(\gamma_{ij}(v),\gamma_{jk}(v))^{-1}\rho(\gamma_{ik}(v))\psi(v),\end{align*}
and finally, since $\gamma_{ij}(v)$, $\gamma_{jk}(v) \in N$
we deduce that $u_\rho(\gamma_{ij}(v),\gamma_{jk}(v)) =1$, giving
\begin{equation}
(g_{ij}g_{jk}\phi)(v)
= \phi(\gamma_{ij}(v),\gamma_{jk}(v),\gamma_k(v))^{-1}(g_{ik}\psi)(v).
\end{equation}
This shows that the Dixmier-Douady class can be described by the \v Cech cocycle 
\begin{align*}
\phi_{ijk} 
&= \phi(\gamma_{ij}(v),\gamma_{jk}(v),\gamma_k(v))^{-1}\\
&= \exp[2\pi if(\gamma_{ij}(v),\gamma_{jk}(v),\gamma_k(v))] \\
&= \exp[2\pi if(\gamma_{i}(v),\gamma_{j}(v),\gamma_k(v))],
\end{align*}
where the antisymmetry of $f$ has been used in the final line.
To find a form describing the de Rham cocycle we note that, since locally 
$\gamma_k(v)$ and $v$ are the same and the differences $\gamma_{ij}$ are in 
$N$,
\begin{align*}df(\gamma_{ij}(v),\gamma_{jk}(v),\gamma_k(v)) 
&= f(\gamma_{ij}(v),\gamma_{jk}(v),d\gamma_k(v))\\ 
&= f(\gamma_{ij}(v),\gamma_{jk}(v),dv)\\ 
&= f(\gamma_{ij}(v),\gamma_j(v),dv) - f(\gamma_{ij}(v),\gamma_k(v),dv),\end{align*}
giving an explicit expression as the difference of one-forms.
Repeating this process twice we arrive at the de Rham form $f(dv,dv,dv)$ 
giving the class $H_0$.
(The antisymmetry of $f$ compensates for the antisymmetry of the exterior 
product to give a non-vanishing answer.)
\end{proof}

The same ideas can now be used to deal with the general case.
However, since we want to use results for the untwisted case, we shall at this 
point restrict ourselves to the case of an abelian group $\sfG$

\begin{theorem}
Let $\sfG$ be abelian and $\alg$ a continuous trace algebra with an action of 
$\sfN \subset \sfG$ by locally projectively unitary automorphisms. 
(This includes the assumption that the restriction of $u$ to $\sfN\times \sfN$ 
takes the constant value $1$).
Both $\uind_\sfN^\sfG(\alg)$ and  $\ind_\sfN^\sfG(\alg)$ are continuous trace 
algebras, and the difference of their Dixmier-Douady invariants 
$\delta(\uind_\sfN^\sfG(\alg)) - \delta(\ind_\sfN^\sfG(\alg))$ is the class 
defined by 
the form $f\in \Omega^3(\sfG,\integer)$ associated to the tricharacter $\phi$:
\begin{equation}
\delta(\uind_\sfN^\sfG(\alg)) - \delta(\ind_\sfN^\sfG(\alg)) = [f].
\end{equation}
\end{theorem}

\begin{proof}
We first note that by Theorem \ref{thDa} 
$u$ is exterior equivalent to $u_\lambda$ which by \eqref{eqqDc}
has trivial restriction to $\sfN\times \sfN$.                                                                                                    
Since $\alg$ has continuous trace, it is locally Morita equivalent to an                                         
algebra of  compact operators, and all our calculations wil be local.                                              
In fact we may cover $\widehat{\alg}$ by open sets $\{ U_\lambda\}$ and find 
Hilbert spaces $\hilb_\lambda$ such that the restriction $\alg^{U_\lambda}$ of 
$\alg$ to $U_\lambda$ is Morita equivalent to 
$C(U_\lambda,\cpt(\hilb_\lambda))$ via some bimodule $Y_\lambda$.
Combining these with the bimodules $X_l$ used in the previous proof we take
$\cx_{(l,\lambda)} = X_l\otimes Y_\lambda$ for the restriction of the algebra 
to $\pi(F_l)\times U_\lambda$.

We may assume that the cover is fine enough that $\alpha_n(a)$ is equivalent 
to $\ad(\rho^\lambda_n)(a)$ for $a \in \alg^{U_\lambda}$, with $\rho^\lambda$
a $\theta^\lambda$-representation.
We now define an equivalence $G^{(m,\mu)}_{(l,\lambda)}$ on the overlap of the 
sets $\pi(F_l)\times U_\lambda$ and $\pi(F_m)\times U_\mu$ by setting
\begin{equation}
(G^{(m,\mu)}_{(l,\lambda)}\psi)(v) = u(\gamma_{lm}(v),\gamma_m(v))^{-1}
h_{\lambda\mu}\rho^\mu(\gamma_{lm}(v))\psi(v),
\end{equation}
where $h_{\lambda\mu}$ describes the equivalences of $Y_\lambda$ and $Y_\mu$,
(which are assumed to satisfy the relationship 
$h_{\lambda\mu}h_{\mu\nu} = \Phi_{\lambda\mu\nu}h_{\lambda\nu}$, 
where $\Phi$ is a \v Cech cocycle describing the Dixmier-Douady class of $\alg$.)
On overlaps the projective representations $\rho^\lambda$ are equivalent in 
the sense that $\rho^\mu(n)h_{\mu\nu} = 
h_{\mu\nu}\kappa_{\mu\nu}(n)\rho^\nu(n)$, for some character $\kappa_{\mu\nu} 
\in \widehat{\sfN}$.
The adjoint actions of $\rho^\mu$ and $\rho^\nu$ are both equivalent to 
$\alpha$.

We now calculate that
\begin{align*}(G^{(m,\mu)}_{(l,\lambda)}G_{(m,\mu)}^{(n,\nu)}\psi)(v) 
&= u(\gamma_{lm}(v),\gamma_m(v))^{-1}h_{\lambda\mu}\rho^\mu(\gamma_{lm}(v))
u(\gamma_{mn}(v),\gamma_n(v))^{-1}h_{\mu\nu}\rho^\nu(\gamma_{mn}(v))\psi(v)\\
&= u(\gamma_{lm}(v),\gamma_m(v))^{-1}
\alpha_{\gamma_{lm}})[u(\gamma_{mn}(v),\gamma_n(v))^{-1}]
h_{\lambda\mu}h_{\mu\nu}\\
&\qquad  \times \kappa_{\mu\nu}(\gamma_{lm})
\rho^\nu(\gamma_{lm}(v))\rho^\nu(\gamma_{mn}(v))\psi(v).\end{align*}
The first two terms combine as before to give
\begin{equation}
u(\gamma_{ln}(v),\gamma_n(v))^{-1}u(\gamma_{lm}(v),\gamma_{mn}(v))^{-1}
= \phi(\gamma_{lm},\gamma_{mn},\gamma_n)^{-1}
u(\gamma_{ln}(v),\gamma_n(v))^{-1},
\end{equation}
whilst the projective representions give 
\begin{equation}
\rho^\nu(\gamma_{lm}(v))\rho^\nu(\gamma_{mn}(v))
= \theta^\nu(\gamma_{lm}(v),\gamma_{mn}(v))\rho^\nu(\gamma_{ln}(v)),
\end{equation}
giving
\begin{align*}(G^{(m,\mu)}_{(l,\lambda)}G_{(m,\mu)}^{(n,\nu)}\psi)(v) &= 
\phi(\gamma_{lm},\gamma_{mn},\gamma_n)^{-1}\kappa_{\mu\nu}(\gamma_{lm})
\theta^\nu(\gamma_{lm}(v),\gamma_{mn}(v))\Phi_{\lambda\mu\nu}\\
&\qquad u(\gamma_{ln}(v),\gamma_n(v))^{-1}h_{\lambda\nu}
\rho^\nu(\gamma_{ln}(v))\psi(v)\\
&= \phi(\gamma_{lm},\gamma_{mn},\gamma_n)^{-1}\kappa_{\mu\nu}(\gamma_{lm})
\theta^\nu(\gamma_{lm}(v),\gamma_{mn}(v))\Phi_{\lambda\mu\nu}
G^{(n,\nu)}_{(l,\lambda)}\psi(v),\end{align*}
from which we deduce the obstruction.
All the factors except the first would be present for $\ind_\sfN^\sfG(\alg)$, so 
that the difference between the Dixmier-Douady obstructions for 
$\uind_\sfN^\sfG(\alg)$ and $\ind_\sfN^\sfG(\alg)$ is just given by 
$\phi^{-1}$, or as forms by $f$.
\end{proof} 

Our formula shows that the obstruction has four contributions: 
the MacLane-Whitehead obstruction $\phi^{-1}$, the Mackey obstruction 
$\theta$, the Phillips-Raeburn obstruction $\kappa$, and the Dixmier-Douady 
obstruction $\Phi$ for $\alg$, corresponding to $H_0$, $H_1$, $H_2$ and $H_3$.

\begin{corollary}\label{cor: ind}
Let $\sfG$ be abelian and $E$ be a principal $\sfG/\sfN$-bundle over $M$ with 
prescribed Dixmier-Douady invariant associated with $H$.
Then there is a $u$-induced algebra with spectrum $E$, and the correct action 
of $\sfG$, having the Dixmier-Douady invariant $H$. This  $u$-induced algebra
is not necessarily unique.\footnote{The non uniqueness reflects the fact that 
there may 
exist several liftings of the action of $\sfN$ on $\alg$ to $\sfG$.  So, 
strictly speaking, $\uind$ does
not define a functor, but $\uind_\sfN^\sfG(\alg)$
just indicates the induced algebra with a prescribed action of $\sfG$.}
\end{corollary}

\begin{proof}
We know from \cite{BEMa, BHMa, MR} that there is an algebra 
$\ind_\sfN^\sfG(\alg)$ 
associated with the principal $\sfG/\sfN$-bundle $E$ and having 
Dixmier-Douady form $H-H_0$. 
This algebra is not necessarily unique, cf. \cite{MR}.
With $f$ as in the Theorem we see that $\uind_\sfN^\sfG(\alg)$ has 
Dixmier-Douady class described by the form $H$.
An alternative approach would be to note that a continuous trace algebra with Dixmier--Douady       
class given by $H$, can be constructed as the tensor product of two continuous trace algebras with  
classes $H-H_0$ and with $H_0$, and then to apply \cite {MR} for the former and our earlier result   
for the class $H_0$.                                                                                                                                               
\end{proof}

Although we have shown how to construct an algebra with a given 
Dixmier-Douady class, it is natural to wonder whether one could also find 
another algebra with twisting given by an ordinary cocycle.
The next result shows that this is not possible.

\begin{theorem}
Every system with the Dixmier-Douady invariant $H$ is described by 
automorphisms whose twisting gives the same tricharacter $\phi$.
\end{theorem}

\begin{proof}
By the general theory we know that any other system must be exterior 
equivalent 
to the $u$-induced system above, that is it is described by automorphisms 
$\lambda_x = \ad(W(x))\beta_x$, 
and $w(x,y) = W(x)\beta_x[W(y)]v(x,y)W(xy)^{-1}$
for some $M\cb$-valued function $W$ on $\sfG$.
Since the twistings are cohomologous they must define the same class $\phi$, 
but more explicitly we calculate that
\begin{align*}\lambda_x[w(y,z)]w(x,yz) 
&= \ad(W(x))\beta_x[W(y)\beta_y[W(z)]v(y,z)W(yz)^{-1}]\\
&\qquad W(x)\beta_x[W(yz)]v(x,yz)W(xyz)^{-1}\\
&= W(x)\beta_x[W(y)]\beta_x\beta_y[W(z)]\beta_x[v(y,z)]v(x,yz)W(xyz)^{-1}\\
&= \phi(x,y,z)W(x)\beta_x[W(y)]\ad(v(x,y))\beta_{xy}[W(z)]v(x,y)v(xy,z)
W(xyz)^{-1}\\
&= \phi(x,y,z)W(x)\beta_x[W(y)]v(x,y)W(xy)^{-1}
W(xy)\beta_{xy}[W(z)]v(xy,z)W(xyz)^{-1}\\
&= \phi(x,y,z)w(x,y)w(xy,z),\end{align*}
showing that the same cocycle $\phi$ arises.
\end{proof}


\section{The twisted crossed product algebra}

In this section we can return to the case of a general group $\sfG$.
We have argued that the dual of a bundle described by $\cb$ with a twisted 
group action should be given by the twisted crossed product.
For the induced algebra $\uind_\sfN^\sfG(\alg)$ its twisted crossed product 
with $\sfG$ 
can be calculated explicitly, and the following result shows that it is a 
generalisation of the twisted compact operators in Section 5.

\begin{theorem}\label{thEightone}
The twisted crossed product $\uind_\sfN^\sfG(\alg)\rtimes_{\beta,v}\sfG$ is 
isomorphic to the $*$-algebra of $\alg$-valued kernels on $\sfG\times \sfG$ 
satisfying
\begin{equation}
K_1(rz,rw) = \phi(r,z,w)^{-1}u(r,z)^{-1}\alpha_r[K_1(z,w)]u(r,w),
\end{equation}
with $K^*(z,w) = K(w,z)^*$ and product
\begin{equation}
(K_1\star K_2)(z,w) = \int_\sfG K_1(z,v)K_2(v,w)\phi(z,v,w)\,dv.
\end{equation}
\end{theorem}

\begin{proof}
The twisted crossed product consists of functions $F_j:\sfG\to \cb$ with 
twisted convolution
\begin{equation}
(F_1*F_2)(x) = \int_\sfG F_1(y)\beta_y[F_2(y^{-1}x)]v(y,y^{-1}x)\,dy.
\end{equation}
Identifying $\cb$ with functions from $\sfG$ to $\alg$, we may give the 
elements of the twisted crossed product a second argument in $\sfG$ and, 
using the explicit form of the induced action and of $v$, get
\begin{equation}
(F_1*F_2)(z,x) 
= \int_\sfG F_1(z,y)\ad(u(z,y))[F_2(zy,y^{-1}x)]
\phi(z,y,y^{-1}x)u(z,y)u(zy,y^{-1}x)u(z,x)^{-1}\,dy,
\end{equation}
which can be rearranged as
\begin{equation}
(F_1*F_2)(z,x)u(z,x) 
= \int_\sfG \phi(z,y,y^{-1}x)F_1(z,y)u(z,y)F_2(zy,y^{-1}x)u(zy,y^{-1}x)
\,dy.
\end{equation}
We now define $K_1(z,zy) = F_1(z,y)u(z,y)$, $K_2(z,zy) = F_2(z,y)u(z,y)$ and 
$(K_1\star K_2)(z,zx) = (F_1*F_2)(z,x)u(z,x)$ to obtain
\begin{equation}
(K_1\star K_2)(z,zx) 
= \int_\sfG K_1(z,zy)K_2(zy,zx)\phi(z,y,y^{-1}x)\,dy.
\end{equation}
Setting $w=zx$, $v=zy$ and exploiting the antisymmetry of $\phi$ the result 
follows.
We readily check that
\begin{equation}
F^*(z,x)u(z,x) = v(x,x^{-1})^*\ad(u(z,x))[F(zx,x^{-1})]^*u(z,x)
= u(zx,x^{-1})^*F(zx,x^{-1})^*,
\end{equation}
from which it follows that $K^*(z,w) = K(w,z)^*$.

The kernels inherit an equivariance condition from the inducing process.
\begin{align*}K_1(rz,rzy) &= F_1(rz,y)u(rz,y)\\
&=\ad(u(r,z))^{-1}\alpha_r[F_1(z,y)]u(rz,y)\\
&= u(r,z)^{-1}\alpha_r[K_1(z,zy)u(z,y)^{-1}]u(r,z)u(rz,y)\\
&= \phi(r,z,y)^{-1}u(r,z)^{-1}\alpha_r[K_1(z,zy)]u(r,zy).\end{align*}
Exploiting the antisymmetry of $\phi$ this gives
\begin{equation}
K_1(rz,rw) = \phi(r,z,w)^{-1}u(r,z)^{-1}\alpha_r[K_1(z,w)]u(r,w).
\end{equation}
We note that since the original product $F_1*F_2$ respected this equivariance 
condition, so does the product on kernels.
The norm on the twisted crossed product can be defined from the left regular 
representation and so agrees with that on kernels.
\end{proof}

\bigskip 
The description of the crossed product algebra in Theorem \ref{thEightone} 
can be made more precise for  a bundle over a point described by $\alg = \cpt(L^2(\sfN))$ 
as in Proposition \ref{thSevenone}
However, this time it is more useful to take the left handed version of the automorphisms, that is 
\begin{equation}
\alpha_r[k]( s,t) = k(r^{-1}s, r^{-1}t) \,,
\end{equation}
and set $(u(s,t)\psi)(r) = \phi(r,s,t)^{-1}\psi(r)$.

\begin{theorem} \label{thHb}
The non-associative torus describing the dual of a bundle over a point is isomorphic to the algebra $A_\phi = \cpt_\phi(L^2(\sfG))\times_{\gamma,u} \sfN$.
\end{theorem}

\begin{proof}
We abbreviate notation for the $\cpt(L^2(\sfN))$-valued kernels by setting 
\begin{equation}
K(z,w):  (s,t) \mapsto K(z,w;s,t),
\end{equation}
so that the equivariance condition can be given explicitly as
\begin{equation}
K(rz,rw;s,t) = \phi(r,z,w)^{-1}\phi(r,z,s)K(z,w;r^{-1}s,r^{-1}t)]\phi(r,w,t)^{-1},
\end{equation}
or, replacing the first two arguments, as
\begin{equation}
K(z,w;s,t) = \phi(r,z,w)^{-1}\phi(r,z,s)K(r^{-1}z,r^{-1}w;r^{-1}s,r^{-1}t)]\phi(r,w,t)^{-1}.
\end{equation}
Taking $r=s$ in this formula we get
\begin{equation}
K(z,w;s,t) = \phi(s,z,w)^{-1}K(s^{-1}z,s^{-1}w;1,s^{-1}t)\phi(w,s,t).
\end{equation}
Now there is an automorphism $\gamma_s$ of the algebra of twisted kernels $\cpt_\phi(L^2(\sfG))$ 
given by
\begin{equation}
\gamma_s[K](z,w) = \phi(s,z,w)^{-1}K(s^{-1}z,s^{-1}w)
\end{equation}
which is just the left-handed version of the automorphism $\theta_s$ of  introduced in Proposition 5.1, and is associated with the multiplier $u(s,t)$ so that
\begin{equation}
K(z,w;s,t) = \gamma_s[K](z,w;1,s^{-1}t)u(s,t)
= \gamma_s[K](z,w;1,s^{-1}t)u(s,s^{-1}t)\,.
\end{equation}

This shows that the kernels can be reconstructed from their values when the third argument is 1, 
and in this case the product formula can be simplified.
The product
\begin{equation}
(K_1\star K_2)(z,w;rt) = \int_{N\times G} K_1(z,v;r,s)K_2(v,w;s,t)\phi(z,v,w)\,dsdv\,,
\end{equation}
reduces to 
\begin{equation}
(K_1\star K_2)(z,w;1,t) = \int_{N\times G} K_1(z,v;1,s)K_2(v,w;s,t)\phi(z,v,w)\,dsdv\,.
\end{equation}
The second kernel can be rewritten using the equivariance condition to give
\begin{equation}
(K_1\star K_2)(z,w;1,t) 
= \int_G\{ \int_N K_1(z,v;1,s)\gamma_s[K_2](v,w;1,s^{-1}t)u(s,s^{-1}t)\,ds\}\phi(z,v,w)\,dv\,.
\end{equation}
Identifying the kernel $K_j$ with the $\cpt_\phi(L^2(\sfG))$-valued function 
$s \mapsto \{(z,w) \mapsto K_j(z,w;1,s)\}$ on $N$, this is just the twisted crossed product of the 
two functions.
In other words we can identify the algebra with  $\cpt_\phi(L^2(\sfG))\times_{\gamma,u} \sfN$.
\end{proof}

\bigskip
In the general case of a bundle over $M$ one needs to take $\alg = C(M,\cpt(L^2(\sfN))$, but 
since the products of functions on $M$ are all taken pointwise, there is no essential change 
in the calculations. 
The only point for caution is that, in those cases where  $\phi$ depends on $M$ 
(and consequently so also do $\gamma$ and $u$), one is really looking at an algebra 
of continuous sections of a bundle over $M$ rather than just functions.

\begin{theorem}
The non-associative torus describing the dual of a bundle over $M$ is isomorphic to 
the algebra $C(M,\cpt_\phi(L^2(\sfG))\rtimes_{\gamma,u} \sfN)$.
\end{theorem}

When $\sfN$ is trivial this gives the twisted algebra of kernels introduced in 
Section 5, though with $\phi$ replaced by its inverse.
More generally it provides a an extension of Green's Generalised Imprimitivity 
Theorem \cite{Gre} to the case of these twisted induced algebras.

In contrast to the twisted crossed product algebra for the dual, the 
correspondence algebra is associative.
Provided that there is no Mackey obstruction the correspondence space is 
associated with the algebra $\uind_\sfN^\sfG(\alg)\rtimes_{\beta,v}\sfN$.

\begin{theorem}
The algebra $\uind_\sfN^\sfG(\alg)\rtimes_{\beta,v}\sfN$ is associative.
\end{theorem}

\begin{proof}
By Propositions 3.1 and 6.4 $\uind_\sfN^\sfG(\alg)\rtimes_{\beta,v}\sfN$ is 
associative if and only if the restriction of $\phi$ to 
$\sfN\times \sfN\times \sfN$  is 1, and we 
have seen that this is a consequence of the integrality of the form $H_0$.
\end{proof}

For non-trivial Mackey obstruction one replaces $\sfN$ by the projection onto 
$\sfG$ of the centre of the central extension defined by the multiplier 
\cite{Ha,ER}, and that algebra is similarly associative.


\section{The dual action}

In this section we shall assume that $\sfG$ is abelian.
Following the development in Section 5, we may define automorphisms of the 
kernels by
\begin{equation}
\tau_x[K](z,w) = \phi(x,z,w)K(x^{-1}z,x^{-1}w),
\end{equation}
and these satisfy 
$\tau_x\tau_y = \ad(\widetilde{u}(x,y))^{-1}\tau_{xy}$, where
\begin{equation}
\ad(\widetilde{u}(x,y)[K])(z,w)=\phi(x,y,z)\phi(x,y,w)^{-1}K(z,w).
\end{equation}
However, we are primarily interested in the case of abelian groups, and 
there is also a more useful action of the dual group $\wh{\sfG}$ on the 
dual algebra.
Its definition is motivated by the action $\wh{\beta}_\xi$ on 
$F \in \uind_\sfN^\sfG(\alg)\rtimes_{\beta,v}\sfG$ given by
$\wh{\beta}_\xi[F](z,y) = \xi(y)F(z,y)$. Rewritten in terms of kernels this 
leads to the following idea.

\begin{proposition}
For $\xi\in\wh{\sfG}$ define  $\wh{\beta}_\xi$ by
\begin{equation}
\wh{\beta}_\xi[K](z,w) = \xi(z^{-1}w)K(z,w).
\end{equation}
Then $\wh{\beta}_\xi$ is an automorphism of 
$\uind_\sfN^\sfG(\alg)\rtimes_{\beta,v}\sfG$,
and $\wh{\beta}_\xi\wh{\beta}_\eta = \wh{\beta}_{\xi\eta}$.
\end{proposition}

\begin{proof}
We must first show that $\wh{\beta}_\xi$ is well-defined, that is that it
preserves the subspace of kernels $K$ satisfying the equivariance condition 
of Theorem 8.1.
In fact we see that
\begin{align*}\wh{\beta}_\xi[K](rz,rw) 
&= \xi(z^{-1}w)K(rz,rw)\\
&= \xi(z^{-1}w)\phi(r,z,w)^{-1}u(r,z)^{-1}\beta_r[K(z,w)]u(r,w)\\
&= \phi(r,z,w)^{-1}u(r,z)^{-1}\beta_r[\wh{\beta}_\xi[K](z,w)]u(r,w),\end{align*}
which gives the required equivariance condition.
It is also an automorphism since
\begin{align*}(\wh{\beta}_\xi[K_1]\star \wh{\beta}_\xi[K_2])(z,w) 
&= \int_\sfG \xi(z^{-1}v)K_1(z,v)\xi(v^{-1}w)K_2(v,w)\phi(z,v,w)^{-1}\,dv\\
&= \int_\sfG \xi(z^{-1}w)K_1(z,v))K_2(v,w)\phi(z,v,w)^{-1}\,dv\\
&= \xi(z^{-1}w)(K_1\star K_2)(v,w).\end{align*}
There is no twisting involved since it is easy to check that 
$\wh{\beta}_\xi\wh{\beta}_\eta = \wh{\beta}_{\xi\eta}$.
\end{proof}

We now form the crossed product of the twisted crossed product algebra with 
$\wh{\sfG}$.
The elements are functions from $\wh{\sfG}$ to the algebra of kernels 
described 
in the last Section, and so may be regarded as $\alg$-valued functions on 
$\sfG\times \sfG\times \wh{\sfG}$, with multiplication
\begin{align*}(\wh{k}_1\star \wh{k}_2)(z,w,\xi) 
&= \int_{\sfG\times\wh{\sfG}} \wh{k}_1(z,v,\eta)\wh{\beta}_\eta[\wh{k}_2]
(v,w,\eta^{-1}\xi)\phi(z,v,w)^{-1}\,d\eta dv\\
&= \int_{\sfG\times\wh{\sfG}} \wh{k}_1(z,v,\eta)\eta(v^{-1}w)
\wh{k}_2(v,w,\eta^{-1}\xi)
\phi(z,v,w)^{-1}\,d\eta dv.\end{align*}

We shall denote the group-theoretic Fourier transform of $\wh{k}$ with respect 
to its third argument by $k$ (which is now a function on 
$\sfG\times \sfG\times \sfG$):
\begin{equation}
k(z,w,x) = \int_{\wh{\sfG}} \wh{k}(z,w,\xi) \xi(x)\,d\xi.
\end{equation}
The multiplication obtained by Fourier transform is (assuming appropriate 
normalisation of the measures)
\begin{align*}(k_1\star k_2)&(z,w,x) 
= \int_{\sfG\times\wh{\sfG}\times\wh{\sfG}} \xi(x) \wh{k}_1(z,v,\eta)
\eta(v^{-1}w)
\wh{k}_2(v,w,\eta^{-1}\xi) \phi(z,v,w)^{-1}\,d\xi d\eta dv\\
&= \int_{\sfG\times\wh{\sfG}\times\wh{\sfG}}  \wh{k}_1(z,v,\eta)\eta(v^{-1}wx)
\wh{k}_2(v,w,\eta^{-1}\xi) (\eta^{-1}\xi)(x)\phi(z,v,w)^{-1}\,d\xi d\eta dv\\
&= \int_\sfG
k_1(z,v,v^{-1}wx)k_2(v,w,x) \phi(z,v,w)^{-1}\,dv.\end{align*}
We now introduce another transformation by
\begin{equation}
\wt{k}(x;z,w) = \phi(x,z,w)^{-1}u(x,z)k(xz,xw,w^{-1})u(x,w)^{-1}.
\end{equation}

\begin{theorem}\label{thm: Morita}
There is an isomorphism
\begin{equation}
(\uind_\sfN^\sfG(\alg)\rtimes_{\beta,v} \sfG)\rtimes_{\wh{\beta}}\wh{G}
\cong \uind_\sfN^\sfG(\alg)\otimes\cpt_{\conj{\phi}}(L^2(\sfG)).
\end{equation}
\end{theorem}

\begin{proof}
One important property which follows from the equivariance condition for the 
$\alg$-valued kernels is that
\begin{align*}\wt{k}&(rx;z,w) 
= \phi(rx,z,w)^{-1}u(rx,z)k(rxz,rxw,w^{-1})u(rx,w)^{-1}\\
&= [\phi(rx,z,w)\phi(r,xz,xw)]^{-1}
u(rx,z)u(r,xz)^{-1}\alpha_r[k(xz,xw,w^{-1})]u(r,xw)u(rx,w)^{-1}\\
&= [\phi(rx,z,w)\phi(r,xz,xw)]^{-1}\phi(r,x,z)^{-1}\phi(r,x,w)\\
&\qquad u(r,x)^{-1}\alpha_r[u(x,z)]\alpha_r[k(xz,xw,w^{-1})]
\alpha_r[u(x,w)]^{-1}u(r,x)\\
&= [\phi(rx,z,w)\phi(r,xz,xw)]^{-1}\phi(r,z,x)\phi(r,x,w)\\
&\qquad \ad(u(r,x))^{-1}\alpha_r[u(x,z)k(xz,xw,w^{-1})u(x,w)^{-1}]\\
&= \ad(u(r,x))^{-1}\alpha_r[\wt{k}(x;z,w)],\end{align*}
so that the kernels $\wt{k}$ satisfy the induced algebra condition with 
respect to $x$, and so can be considered elements of the algebra induced from 
$\sfN$ to $\sfG$ by the $\alg$-valued kernels.

Next we consider the product on these functions
\begin{align*}(\wt{k}_1\star \wt{k}_2)&(x;z,w) 
= \phi(x,z,w)^{-1}u(x,z)(k_1\star k_2)(xz,xw,w^{-1})u(x,w)^{-1} \\
&= \phi(x,z,w)^{-1}\int_\sfG u(x,z)k_1(xz,xv,v^{-1})k_2(xv,xw,w^{-1})
u(x,w)^{-1} \phi(xz,xv,xw)^{-1}\,dv\\
&= \int_\sfG \wt{k}_1(x;z,v)\wt{k}_2(x;v,w) \phi(z,v,w)^{-1}\,dv.\end{align*}
Thus we have the pointwise product with respect to $x$, with the twisted 
multiplication of $\cpt_{\conj{\phi}}(L^2(\sfG))$ in the fibres.
\end{proof}

We shall show in the final section that the twisted compact operators
are in a certain sense Morita equivalent to the ordinary compact
operators, so that this result provides a very precise analogue of the
normal duality theorem.

To complete the argument we really need to know that the double dual action 
$\wh{\wh{\beta}}$ is equivalent to the original 
$\beta$ to within the action on 
the twisted compact operators. 
The double dual actin is defined on 
$(\uind_\sfN^\sfG(\alg)\rtimes_{\beta,v} \sfG)\rtimes_{\wh{\beta}}\wh{\sfG}$
by the same procedure used to obtain the dual action, that is multiplication 
by the pairing of the $\wh{\sfG}$ variable and the group element:
\begin{equation}
(\wh{\wh{\beta}}_g[\wh{k}](z,w,\xi) = \xi(g)\wh{k}(z,w,\xi).
\end{equation}

\begin{theorem}
The double dual action of $\sfG$ can be written as
\begin{equation}
(\wh{\wh{\beta}}_g[\wt{k}])(x;z,w) 
= \tau_g[(v(g,g^{-1}z)^{-1}\beta_g[\wt{k}]v(g,g^{-1}w)(x;z,w)].
\end{equation}
\end{theorem}

\begin{proof}
Fourier transforming the definition of the action we get
\begin{equation}
(\wh{\wh{\beta}}_g[k](z,w,x) = \int_{\wh{\sfG}}\xi(g)\xi(x)\wh{k}(z,w,\xi)
= k(z,w,xg).
\end{equation}
Using the same notation for the equivalent action on $\wt{k}$
\begin{align*}(\wh{\wh{\beta}}_g[\wt{k}])(x;z,w) 
&= \phi(x,z,w)^{-1}u(x,z)(\wh{\wh{\beta}}_g[k])(xz,xw,w^{-1})u(x,w)^{-1}\\
&= \phi(x,z,w)^{-1}u(x,z)k(xz,xw,w^{-1}g)u(x,w)^{-1}\\
&= \phi(xg,g^{-1}z,g^{-1}w)\phi(x,z,w)^{-1}u(x,z)u(xg,g^{-1}z)^{-1}\\
&\qquad \wt{k}(xg;g^{-1}z,g^{-1}w)u(xg,g^{-1}w)u(x,w)^{-1}.\end{align*}
In terms of the induced twisting we now we have 
\begin{equation}
u(x,z)u(xg,g^{-1}z)^{-1} = 
\phi(x,g,g^{-1}z)^{-1}v(g,g^{-1}z)(x)^{-1}u(x,g),
\end{equation}
and substituting this (and the analogous expression in $w$), we arrive at
\begin{align*}(\wh{\wh{\beta}}_g[\wt{k}])(x;z,w) 
&= \phi(xg,g^{-1}z,g^{-1}w)\phi(x,z,w)^{-1}\phi(x,g,g^{-1}z)^{-1}
\phi(x,g,g^{-1}w)v(g,g^{-1}z)(x)^{-1}\\
&\qquad u(x,g)\wt{k}(xg;g^{-1}z,g^{-1}w)
u(x,g)^{-1}v(g,g^{-1}w)(x)\\
&= \phi(g,z,w)v(g,g^{-1}z)(x)^{-1}\ad(u(x,g))[\wt{k}(xg;g^{-1}z,g^{-1}w)]
v(g,g^{-1}w)(x)\\
&= \phi(g,z,w)v(g,g^{-1}z)(x)^{-1}(\beta_g[\wt{k}](x;g^{-1}z,g^{-1}w)]
v(g,g^{-1}w)(x).\end{align*}
This can be rewritten in terms of the twisted action $\tau_g$ on kernels 
and the adjoint action of $v$ in the form
\begin{equation}
(\wh{\wh{\beta}}_g[\wt{k}])(x;z,w) 
= \tau_g[(v(g,g^{-1}z)^{-1}\beta_g[\wt{k}]v(g,g^{-1}w))(x;z,w)],
\end{equation}
showing that to within an inner automorphism of the kernels one has
$\beta_g\otimes \tau_g$, and up to an action on the twisted kernels one 
recovers $\beta_g$.
\end{proof}

In the application to principal $\torus^n$-bundles one takes $\sfG = \real^n$.
As this is contractible, Proposition 5.2 shows that 
$\cpt_{\conj{\phi}}(L^2(\sfG))$ is a deformation of 
$\cpt(L^2(\sfG))$, so that this 
is a close substitute for the usual duality theorem.


\section{Applications to T-duality}

In this section, we apply the mathematical results of the earlier sections
to determine the T-dual of principal torus bundles with general H-flux,
thus generalizing earlier results in \cite{BEMa, BEMb, BHMa, MR, MRb}. 
Let  $\sfG=\mathbb R^\ell$, and $\sfN=\mathbb Z^\ell$ in the setup of Corollary \ref{cor: ind}.

Let $E\to M$ be a principal $\mathbb T^\ell$-bundle, and $H\in H^3(E)$ 
be a an H-flux on $E$.\footnote{The conclusions in this Section are valid for
integral classes $H\in H^3(E,\ZZ)$ as well since the component $H_0$ does not
have carry torsion, and the remainder of the arguments is based on 
the results of \cite{MR,MRb}, which also hold for torsion $H$.  For simplicity 
we state the results for differential forms only.}
Then we can identify $H=(H_3,H_2,H_1,H_0)$, where 
$H_p \in \Omega^p(M, \wedge^{3-p}\ \wh{\ft})$, under the isomorphism 
\eqref{eqAa}, and closed under $D$. Then by the results in \cite{MR, MRb}, there is 
a continuous trace $C^*$-algebra $\ind_\sfN^\sfG(\alg)$
with spectrum equal to $E$ and with Dixmier-Douady invariant
equal to $(H_3,H_2,H_1,0)$, which has an action of $\sfG$ that covers
the given action of $\sfG$ on $E$. This action is not necessarily unique. 
Then by Corollary \ref{cor: ind}, we know that there is another 
continuous trace $C^*$-algebra $\uind_\sfN^\sfG(\alg)$
with spectrum equal to $E$ and with Dixmier-Douady invariant
equal to $H=(H_3,H_2,H_1,H_0)$, which has a twisted action of $\sfG$ that covers
the given action of $\sfG$ on $E$.  Our main definition in this section is:

\begin{definition}
The twisted crossed product 
$\uind_\sfN^\sfG(\alg)\rtimes_{\beta,v}\sfG$ is defined
to be the T-dual to the principal $\TT^\ell$-bundle $E$ with H-flux $H$.
\end{definition}

We justify this definition as follows. Firstly, the T-dual of $\uind_\sfN^\sfG(\alg)\rtimes_{\beta,v}\sfG$
is the crossed product $(\uind_\sfN^\sfG(\alg)\rtimes_{\beta,v} \sfG)\rtimes_{\wh\beta} \wh{\sfG}$, which by 
the twisted Takai duality Theorem \ref{thm: Morita} is isomorphic to $\uind_\sfN^\sfG(\alg)\otimes\cpt_\phi(L^2(\sfG))$.
That is, the T-dual of $\uind_\sfN^\sfG(\alg)\rtimes_{\beta,v}\sfG$ is Morita equivalent 
to the continuous trace algebra $\uind_\sfN^\sfG(\alg)$, so that T-duality applied twice
returns us to where we started, up to Morita equivalence. 
In the special case when $H=H_0$, the fibre of this bundle 
over the point $z\in M$ is equal to the nonassociative torus
$A_\phi $  of rank $\ell$ with tricharacter 
$\phi$ corresponding to $H_0$ (see Theorem \ref{thHb}).
In the general case, but when $H_0$ is zero, the fibre is a stabilized noncommutative 
torus with
invariant $H_1$, \cite{MR, MRb}, and when $H_0=0$ and $H_1=0$, then the fibre is the stabilized 
algebra  of continuous functions on a torus, \cite{BEMa, BEMb, BHMa}. Thus we have the following theorem.

\begin{theorem}[{\bf T-duality for principal torus bundles}]\label{thm:tduality}
Let $E\to M$ be a principal $\mathbb T^\ell$-bundle over $M$, and $H\in H^3(E, \mathbb Z)$ 
be a an H-flux on $E$. Then $H=(H_3,H_2,H_1,H_0)$, where 
$H_p \in \Omega^p(M, \wedge^{3-p} \wh{\ft})$. 
Let $c_1(E) \in H^2(M, \ft)$ denote the first Chern class of $E$, which determines $E$
up to isomorphism.

Then:

\begin{enumerate}

\item If $H_0=0$ and $H_1=0$, then there is a uniquely determined T-dual $\widehat E$
which is a principal $\mathbb T^\ell$-bundle over $M$  first Chern class $c_1(\widehat E) = H_2 \in H^2(M, \wh{\ft})$.
$\widehat E$ has a  T-dual H-flux $\widehat H = (\widehat  H_3, \widehat H_2,0,0)$ given by 
$\widehat H_3 = H_3$ and $\widehat H_2= c_1(E)$. T-duality is neatly encapsulated in the 
commutative diagram,
\begin{equation} \label{correspondenceb}
\xymatrix @=5pc @ur { E \ar[d]_{\pi} & 
E\times_M  \widehat E \ar[d]_{\hat p} \ar[l]^{p} \\ M & \widehat E\ar[l]^{\hat \pi}}
\end{equation}

\item If $H_0=0$ and $H_1\ne 0$, then the T-dual is a continuous field of 
(stabilized) noncommutative tori 
$A_f$ over $M$, where the fiber over the point $z\in M$ is equal to the 
rank $k$ noncommutative torus $A_{f(z)}$
(see Figure 1 below).
Here $f : M \to \TT^{\binom \ell 2}$ is a continuous map representing $H_1\in H^1(M,\wedge^2\,
\wh{\ft}) \cong [M, \TT^{\binom \ell 2}]$.
This map is not unique, but the nonuniqueness does not affect its K-theory.

\begin{figure}[ht]
\includegraphics[height=1.5in]{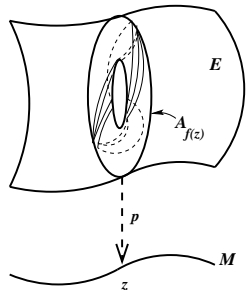}\\
\caption{In the diagram, the fiber over $z\in M$ 
is the noncommutative torus $A_{f(z)}$,
which is represented by a foliated torus, with foliation 
angle equal to $f(z)$.}
\label{fig:folbundle}
\end{figure}

\item If $H_0 \ne 0$ and if $H=H_0$, then the T-dual is a bundle of nonassociative tori $A_\phi$
(cf.\ Theorem \ref{thHb}) over $M$, where $\phi$ is the tricharacter associated to $H_0$. For general
$H$, the T-dual is a continuous field of algebras that contains both the noncommutative torus
and the nonassociative torus, and moreover, the T-dual is not unique, but the
nonuniqueness occurs exactly as in part {\rm (2)} above.
\end{enumerate}

\end{theorem}

Part (1) was proved in \cite{BEMa, BEMb} when $\ell=1$ and in \cite{BHMa} for general $\ell$.

Part (2) was proved in \cite{MR} when $\ell=2$ and in \cite{MRb} for general $\ell$.

Part (3) is what has been proved in this paper. 

\bigskip

A particular, but important case of Theorem \ref{thm:tduality} above is the following.\\

\begin{enumerate}
\item The T-dual of  the torus $\TT^3$ with no background flux is 
the dual torus $\widehat \TT^3$. This remains true if the background flux 
is topologically trivial.\\

\item $(\TT^3, k\; dx \wedge dy \wedge dz)$ considered as a trivial circle bundle over 
$\TT^2$. The T-dual of  $(\TT^3, k\; dx \wedge dy \wedge dz)$ is the nilmanifold
$(H_\RR/H_\ZZ, 0)$, where $H_\RR$ is the 3 dimensional Heisenberg group and
$H_\ZZ$  the lattice in it defined by
 \begin{equation}
H_{\mathbb Z} = \left\{   \begin{pmatrix} 1 & x & \frac{1}{k}z \\
0 & 1 & y\\
0 & 0 &       1
\end{pmatrix} : x, y, z \in \mathbb Z \right\}.
\end{equation}
. \\

\item $(\TT^3, k\; dx \wedge dy \wedge dz)$ considered as a trivial $\TT^2$-bundle over 
$\TT$. The T-dual of  $(\TT^3, k\; dx \wedge dy \wedge dz)$ is a continuous field 
of stabilized noncommutative tori, $C^*(H_\ZZ) \otimes \mathcal K$,
since $\displaystyle\int_{\TT^2}  k\; dx \wedge dy \wedge dz \ne 0$.\\

\item  $(\TT^3, k\; dx \wedge dy \wedge dz)$ considered as a trivial $\TT^3$-bundle over 
a point. The T-dual of  $(\TT^3, k\; dx \wedge dy \wedge dz)$ is a nonassociative
torus, $A_\phi$ (cf.\ Theorem \ref{thHb}), where $\phi$ is the tricharacter associated to 
$ k\; dx \wedge dy \wedge dz $, since $\displaystyle\int_{\TT^3}  k\; dx \wedge dy \wedge dz\ne 0$.\\
\end{enumerate}

We end with some speculations and open problems related to the results of the paper. 
One expects to be 
able to define K-theory for the special nonassociative algebras that are considered
in this paper. These are of the form $\alg\rtimes_{\beta,v}\sfG$, where $\alg$ is a $C^*$-algebra
admitting a twisted action of the Abelian group $\sfG = \mathbb  R^\ell$.  Moreover, there should 
be an analogue of Connes Thom isomorphism theorem in K-theory, showing that the K-theories 
of $\alg$ and $\alg\rtimes_{\beta,v}\sfG$ are naturally isomorphic. This would then give further 
evidence that our definition of the T-dual of a principal torus bundle with H-flux is indeed
correct. Finally, the K-theory of our special nonassociative algebras should be Morita 
invariant in our context, namely invariant under tensor product with twisted compact operators.
Then the twisted Takai duality Theorem \ref{thm: Morita}
would prove that T-duality applied twice returns us to
the torus bundle with  H-flux that we started out with.

It remains to also determine the topological invariants of continuous fields of noncommutative
tori and bundles of nonassociative tori as in the paper. This would then enable one to give 
a more symmetric characterization the T-dual, similar to part (1) of the Theorem above. 
We have an explicit conjecture for this, the explanation for which is in \cite{BHMb}, 
namely, for the continuous field of noncommutative tori $A_f$, there should 
be a ``Chern class'' invariant 
$c_1(A_f) = (H_2,H_1,0)$ satisfying $d H_2 + c_1(E)\wedge H_1 = 0$. 
In this case, we can add the following to part (2) of the Theorem above.

The T-dual  $A_f$ is classified by its Chern class invariant 
$c_1(A_f) = (H_2,H_1,0)$ satisfying $d H_2 + c_1(E)\wedge H_1 = 0$ and $dH_1=0$
and has T-dual H-flux $\widehat H = (\widehat  H_3, \widehat H_2,\widehat H_1,0)$, 
given by $\widehat H_3 = H_3$, $\widehat H_2= c_1(E)$ and $\widehat H_1= 0$.

Similarly, for the bundle of nonassociative tori $A_\phi$ with tricharacter $\phi$ associated
to $H_0$, there should also be a ``Chern class'' invariant 
$c_1(A_\phi) = (H_2,H_1,H_0)$ satisfying $d H_2 + c_1(E)\wedge 
H_1 = 0$ and $dH_1 + c_1(E)\wedge H_0= 0$. 
In this case, we can add the following to part (3) of the Theorem above.

The T-dual  $A_\phi$ is classified by its Chern class invariant 
$c_1(A_\phi) = (H_2,H_1,H_0)$ satisfying $d H_2 + c_1(E)\wedge H_1 = 0$, $dH_1 
+ c_1(E)\wedge H_0= 0$
and $dH_0=0$. It has T-dual H-flux $\widehat H = (\widehat  H_3, \widehat H_2,
 \widehat H_1,\widehat H_0)$ given by $\widehat H_3 = H_3$, $\widehat H_2= c_1(E)$, 
$\widehat H_1= 0$ and $\widehat H_0 = 0$.
 
 What also remains to be done is T-duality for nonabelian principal bundles, where some 
 of the ideas of this paper and \cite{BHMb} apply.

\section{Non-associative algebras and monoidal categories -- An outlook}

Although the non-associativity of the crossed product algebra appears to
present a serious amendment to the notion of duality, that is not really
the case.
The fact that the same obstruction $\phi$ appears throughout is a signal
that
one should rather work in the monoidal category of $C_0(\sfG)$-modules in
which the isomorphism $\Phi: (U\otimes V)\otimes W \to U\otimes (V\otimes
W)$
is given by the action of
$\conj{\phi} \in C(\sfG\times \sfG\times \sfG)$, the multiplier algebra of
$C_0(\sfG)\otimes C_0(\sfG)\otimes C_0(\sfG)$, \cite{McL1, CP}.
The cocycle identity for $\phi$ is equivalent to commutativity of the
fundamental pentagonal diagram which ensures that all higher associators
are
consistent.
By Fourier transforming we could identify the category as $\wh{\sfG}$ modules
rather than $C(\sfG)$-modules, which fits more directly into the framework
of
the duality theorem.
The identity object is the trivial $\wh{\sfG}$-module 1 on $\complex$,
which
certainly has the property that, for any $\wh{\sfG}$-module $U$,
$U\otimes1$ and $1\otimes U$ are naturally isomorphic to $U$.
This is equivalent on $C(\sfG)$ to evaluating a function at the identity.
Because $\phi$ vanishes when an argument is set equal to the identity, the
two
obvious maps from $U\otimes (1\otimes V) = \Phi[(U\otimes 1)\otimes V]$ to
$U\otimes V$ are consistent.

An algebra $\alg$ is a  monoid in this category, and the identification
$\Phi$
automatically takes care of the associativity.
We can also define a left $\alg$-module $M$ if one has a morphism
$\alg \otimes M \to M$.

For example,  the algebra $\cpt_{\conj{\phi}}(L^2(\sfG))$ of twisted
compact
operators has the $\wh{\sfG}$ action
$$(\xi.K)(x,y) = \xi(xy^{-1})K(x,y)$$
and $L^2(\sfG)$, which has the
$\wh{\sfG}$-action $(\xi.\psi)(x) = \xi(x)\psi(x)$, is a module with
$K\otimes \psi \mapsto K*\psi$ where
$$(K*\psi)(x) = \int_\sfG K(x,z)\psi(z)\,dz.$$
The $\wh{\sfG}$ actions are compatible since
$$((\xi.K)*(\xi.\psi))(x) = \int_\sfG \xi(xz^{-1})K(x,z)\xi(z)\psi(z)\,dz
= (\xi(x)(K*\psi))(x).$$
Then
$$(K_1*(K_2*\psi))(x) = \int_\sfG K_1(x,y)K_2(y,z)\psi(z)\,dydz,$$
and the alternate bracketing $(K_1*K_2)*\psi$ must be computed as the
image of
$\Phi(K_1\otimes(K_2\otimes\psi))$, giving
$$((K_1*K_2)*\psi)(x)
= \int_\sfG \phi(x,y,z)^{-1}K_1(x,y)K_2(y,z)\psi(z)\,dydz,$$
consistent with the multiplication law on the twisted kernels.

One can alternately work with the $C_0(\sfG)$ action rather than
$\wh{\sfG}$
but then the action on kernels requires a use of the coproduct
$(\Delta f)(x,y) = f(xy)$, so that $(f.K)(x,y) = f(xy^{-1})K(x,y)$.

One can similarly define right $\alg$-modules, and also bimodules for two
algebras $\alg_1$ and $\alg_2$.
It is also possible to look at $\alg_1$-$\alg_2$-bimodules $X$ which have
an
action of a product group $\wh{\sfG}_1\times\wh{\sfG}_2$ with maps
$\Phi_1$ and $\Phi_2$ defining the associativity properties, and are in
the
category of $\wh{\sfG}_1$-modules as left $\alg_1$-modules and in the
category
of $\wh{\sfG}_2$-modules as right $\alg_2$-modules.
Such a bimodule can be used to set up a Morita equivalence between
left $\alg_2$-modules and left $\alg_1$-modules, by mapping a left
$\alg_2$-module $M$ to the quotient of $X\otimes V$ by the equivalence
relation
$(x.b)\otimes \psi \sim \Phi_2(x\otimes(b.\psi))$, for $x\in X$,
$b\in \alg_2$ and $\psi\in M$.
This allows us to define Morita equivalence between algebras with
different
kinds of associativity.
In particular, if we take $\alg_1 = \cpt_{\conj{\phi}}(L^2(\sfG))$,
$\alg_2 = \cpt(L^2(\sfG))$, with $X = \cpt(L^2(\sfG))$, equipped with the
usual right multiplication action of $\alg_2$ and the left multiplication
action of $\alg_1$ defined above, then we have Morita equivalence between
the twisted and untwisted algebras.

Clearly this is only an outline of some of the ideas arising out of this
new perspective on non-associativity, and we shall explore these in more
detail in the sequel to this paper.
Since posting this paper on the arXives \cite{AM} has come to our attention, which also  
investigates some non-associative algebras albeit in a rather different context.              

\bigskip

\noindent{\bf Acknowledgements:} We would like to thank Edwin
Beggs, Alan Carey and Ulrike Tillmann for useful comments.
KCH would like to thank the University of Adelaide for hospitality 
during the initial stages of the project.
PB and VM were financially supported by the Australian Research Council.



\begin{thebibliography}{99}

\bibitem{AM}                                                                                                                                      
S.E.Akrami and S. Majid,                                                                                                                
{\it Braided cyclic cocycles and non-associative geometry}                                                    
J. Math. Phys. {\bf 45} (2004),
[{\tt arXiv:math.QA/0406005}].                                                                                                        

\bibitem{BEMa} 
P. Bouwknegt, J. Evslin, and V. Mathai,  
{\it T-duality: Topology change from H-flux}, 
Commun. Math. Phys. {\bf 249} (2004) 383-415, 
[{\tt arXiv:hep-th/0306062}].

\bibitem{BEMb} 
P. Bouwknegt, J. Evslin, and V. Mathai,  
{\it On the topology and H-flux of T-dual manifolds}, 
Phys. Rev. Lett. {\bf 92} (2004) 181601, 
[{\tt arXiv:hep-th/0312052}].

\bibitem{BHMa} 
P. Bouwknegt, K.C. Hannabuss, and V. Mathai, 
{\it T-duality for principal torus bundles}, 
J. High Energy Phys. {\bf 03}  (2004) 018, 
[{\tt arXiv:hep-th/0312284}].

\bibitem{BHMb} 
P. Bouwknegt, K.C. Hannabuss, and V. Mathai, 
{\it T-duality for principal torus bundles and dimensionally reduced Gysin 
sequences}, 
in preparation.

\bibitem{Bro} 
K.S.  Brown, 
{\it Cohomology of groups}, 
Springer Verlag, New York--Berlin, 1982.

\bibitem{BS} 
R.C. Busby and H.A. Smith,
{\it Representations of twisted group algebras}, 
Trans. Amer. Math. Soc. {\bf 149} (1970) 503-537.

\bibitem{Car} 
A.L. Carey, 
{\it The origin of three-cocycles in quantum field theory}, 
Phys. Lett.  {\bf B194} (1987) 267-270.

\bibitem{CHMM} 
A.L. Carey, K.C. Hannabuss, V. Mathai, and P. McCann,
{\it The quantum Hall effect in hyperbolic space}, 
Commun. Math. Phys. {\bf 190} (1998) 629-673.

\bibitem{CM} 
A.L. Carey and J. Mickelsson, 
{\it The universal gerbe, Dixmier--Douady class, and gauge theory}, 
Lett. Math. Phys. {\bf 59} (2002) 47-60.

\bibitem{CKRW} 
D. Crocker, A. Kumjian, I. Raeburn, and D. Williams,
{\it An equivariant Brauer group and actions of groups on C$^*$ algebras},
J. Func. Anal. {\bf 146} (1997) 151-184.

\bibitem{CP} 
V. Chari and A. Pressley,
{\it A guide to quantum groups},
Cambridge University Press, Cambridge, 1994.

\bibitem{CS} 
L. Cornalba and R. Schiappa, 
{\it Nonassociative star product deformations for $D$-brane worldvolumes 
in curved backgrounds},
Commun. Math. Phys. {\bf 225} (2002) 33-66,
[{\tt arXiv:hep-th/0101219}].

\bibitem{ER} 
S. Echterhoff and J. Rosenberg,  
{\it Fine structure of the Mackey machine for actions of abelian groups 
with constant Mackey obstruction}, 
Pacific J. Math. {\bf 170}  (1995) 17-52.

\bibitem{EM1} 
S. Eilenberg and S. MacLane, 
{\it Cohomology theory in  abstract groups II}, 
Ann. Math. {\bf 48} (1947) 326-341.

\bibitem{EM2} 
S. Eilenberg and S. MacLane, 
{\it Algebraic cohomology and loops},  
Duke Math. J. {\bf 14} (1947) 435-463.

\bibitem{Gre} 
P. Green,
{\it The structure of imprimitivity algebras}, 
J. Funct. Anal. {\bf 36} (1980) 88-104.

\bibitem{GHV}
W.~Greub, S.~Halperin and R.~Vanstone,
{\it Connections, curvature, and cohomology}, Vols II and III,
Academic Press, New York, 1973.

\bibitem{Ha}  
K.C. Hannabuss, 
{\it Representations of nilpotent locally compact groups} , 
J. Funct. Anal.  {\bf 34}  (1979) 146-165.

\bibitem{Ho} 
D.F.  Holt, 
{\it Cohomology groups $H^n(G,M)$},  
J.  Alg.  {\bf 60} (1979) 307-320.

\bibitem{Hu} 
J.  Huebschmann, 
{\it Crossed $n$-fold extensions of groups and cohomology}, 
Comm. Math. Helv. {\bf 55} (1980) 302-313.

\bibitem{Jac} 
R. Jackiw, 
{\it Three-cocycles in mathematics and physics}, 
Phys. Rev.  Lett. {\bf 54} (1985) 159-162.

\bibitem{Kle} 
A.A. Kleppner,  
{\it Multipliers on abelian groups}, 
Math. Ann. {\bf 158} (1965) 11-34.

\bibitem{Lep} 
H. Leptin, 
{\it Verallgemeinerte $L^1$-Algebren}, 
Math. Ann. {\bf 159} (1965) 51-76.

\bibitem{McL1} 
S. MacLane, 
{\it Categories for the working mathematician},  
Springer Verlag, New York, 1971.

\bibitem{McL} 
S. MacLane, 
{\it Historical Note},  
J. Alg.  {\bf 60} (1979) 319-320.

\bibitem{MW} 
S. MacLane and J.H.C. Whitehead, 
{\it On the 3-type of a complex}, 
Proc.Nat.Acad. Sci. U.S.A. {\bf 30} (1950) 41-48.

\bibitem{MR} 
V. Mathai and J. Rosenberg,  
{\it T-duality for torus bundles with H-fluxes via noncommutative topology} , 
Commun. Math. Phys. {\bf } (2004), 
[{\tt arXiv:hep-th/0401168}].

\bibitem{MRb} 
V. Mathai and J. Rosenberg,  
{\it On mysteriously missing T-duals, H-flux and the T-duality group},
[{\tt arXiv:hep-th/0409073}].  

\bibitem{PR} 
J.A. Packer and I. Raeburn, 
{\it Twisted crossed products of C$^*$-algebras}, 
Math. Proc. Camb. Phil. Soc. {\bf 106} (1989) 293-311.

\bibitem{Qui} 
J.C. Quigg,
{\it Duality for reduced twisted crossed products of $C^*$-algebras}, 
Indiana Univ. Math. J. {\bf 35} (1986) 549-571.

\bibitem{RSW} 
I. Raeburn, A. Sims and D. Williams, 
{\it Twisted actions and obstructions in group cohomology},  in {\it $C^*$-algebras},
ed. J. Cuntz and S. Echterhoff, Springer Verlag, Berlin, 2000.

\bibitem{RW} 
I. Raeburn, and D. Williams, 
{\it Morita equivalence and continuous trace $C^*$ algebras}, 
Mathematical Surveys and Monographs of the A.M.S.
{\bf 60}, Amer. Math. Soc. Providence 1998.

\bibitem{Rat} 
J.G. Ratcliffe, 
{\it Crossed extensions}, 
Trans. Amer. Math. Soc. {\bf 257} (1980) 73-89.

\bibitem{Whi} 
J.H.C. Whitehead, 
{\it Combinatorial homotopy II},  
Bull. Amer. Math. Soc. {\bf 55} (1949) 453-496.



\end{thebibliography}
\end{document}